\PassOptionsToPackage{hyphens}{`}
\documentclass[conference]{IEEEtran}
\IEEEoverridecommandlockouts

\newif\ifcomments
\commentstrue

\newif\ifdraft
\draftfalse

\newif\ifauthors
\authorstrue

\usepackage{amssymb}
\usepackage{array}
\usepackage{mathtools}
\usepackage{amsmath}
\usepackage{wasysym}
\usepackage{makecell}
\usepackage{pifont}
\usepackage{scalerel}

\usepackage{rotating}

\newcommand{\cheatspace}{\vspace{-0.0cm}}

\usepackage{booktabs} % For formal tables
\usepackage{epsfig,endnotes}

\usepackage{epstopdf}
\usepackage{xcolor}

\let\labelindent\relax
\usepackage{enumitem}
\usepackage{float}
\usepackage{xspace}
\usepackage{caption}
\usepackage{multirow}
\usepackage{listings}
\usepackage{tikz}
\usepackage{subfig}

\usepackage{color, colortbl}
\definecolor{Gray}{gray}{0.9}

\ifdraft
  	\usepackage[printwatermark]{xwatermark}
   	\newwatermark[allpages,color=red!50,angle=45,scale=3,xpos=0,ypos=0]{DRAFT}
\else
	\usepackage{url}
\fi

\ifcomments
\newcommand{\note}[1]{\textbf{\textcolor{blue}{[#1]}}}
\newcommand{\jason}[1]{\textbf{\textcolor{red}{[#1]}}}
\newcommand{\nagad}[1]{\textbf{\textcolor{orange}{[#1]}}}
\newcommand{\yagad}[1]{\textbf{\textcolor{purple}{[#1]}}}
\else
    \newcommand{\note}[1]{}
    \newcommand{\jason}[1]{} 
    \newcommand{\nagad}[1]{} 
    \newcommand{\yagad}[1]{}
\fi

\definecolor{apink}{HTML}{FF6666}
\definecolor{agray}{HTML}{CCCCCC}

\colorlet{punct}{red!60!black}
\definecolor{background}{HTML}{EEEEEE}
\definecolor{Gray}{gray}{0.95}
\definecolor{delim}{RGB}{20,105,176}
\colorlet{numb}{magenta!60!black}
\definecolor{eminence}{RGB}{108,48,130}

\lstdefinelanguage{bash}{
    basicstyle=\scriptsize\ttfamily,
    showstringspaces=false,
    keywords={adb,shell},
    keywordstyle=\color{blue},
    breaklines=true,
    morestring = [s]{<}{>},
    stringstyle=\color{red},
    float=t,
    frame=single,
    breakatwhitespace=true,
    captionpos=b,
    comment=[l]{//},
    commentstyle=\itshape\color{green!40!black},
    stringstyle=\color{orange},
    backgroundcolor=\color{background},
    emph={tap,swipe,keyevent,screencap},
    emphstyle={\color{eminence}},
    literate=
     *{0}{{{\color{numb}0}}}{1}
      {1}{{{\color{numb}1}}}{1}
      {2}{{{\color{numb}2}}}{1}
      {3}{{{\color{numb}3}}}{1}
      {4}{{{\color{numb}4}}}{1}
      {5}{{{\color{numb}5}}}{1}
      {6}{{{\color{numb}6}}}{1}
      {7}{{{\color{numb}7}}}{1}
      {8}{{{\color{numb}8}}}{1}
      {9}{{{\color{numb}9}}}{1}
      {X}{{{\color{numb}X}}}{1}
      {Y}{{{\color{numb}Y}}}{1}
      {:}{{{\color{punct}{:}}}}{1}
      {,}{{{\color{punct}{,}}}}{1}
      {\{}{{{\color{delim}{\{}}}}{1}
      {\}}{{{\color{delim}{\}}}}}{1}
      {[}{{{\color{delim}{[}}}}{1}
      {]}{{{\color{delim}{]}}}}{1},
}

\begin{document}

\newcommand{\system}{RF-DCN\xspace}

\ifauthors
\author{
\IEEEauthorblockN{Ioannis Agadakos\IEEEauthorrefmark{1}\thanks{\IEEEauthorrefmark{1} Equal Contributions}}
\IEEEauthorblockA{\small{SRI International}\\
Email: ioannis.agadakos@sri.com\\
}\and
\IEEEauthorblockN{Nikolaos Agadakos\IEEEauthorrefmark{1}\IEEEauthorrefmark{2} }
\IEEEauthorblockA{\small{University of Illinois at Chicago}\\
Email: nagad2@uic.edu }
\and
\IEEEauthorblockN{Jason Polakis}
\IEEEauthorblockA{\small{University of Illinois at Chicago}\\
Email: polakis@uic.edu }
\and
\IEEEauthorblockN{Mohamed R. Amer\IEEEauthorrefmark{2}}\thanks{\IEEEauthorrefmark{2} This work was done while working at SRI International.}
\IEEEauthorblockA{\small{Robust.AI}\\
Email: mohamed@robust.ai }
}
\else
\fi

\title{Deep Complex Networks for Protocol-Agnostic Radio Frequency Device Fingerprinting in the Wild}

\maketitle
\begin{abstract}
Researchers have demonstrated various techniques for fingerprinting  and identifying devices. Previous approaches have identified devices
from their network traffic or transmitted signals 
while relying on software or operating system specific artifacts (e.g.,
predictability of protocol header fields) or characteristics of the
underlying protocol (e.g., central frequency offset).
As these constraints can be a hindrance in real-world settings, 
we introduce a practical, generalizable approach that offers significant operational value
for a variety of scenarios, including as an additional factor of authentication for preventing impersonation attacks.
Our goal is to identify artifacts in
transmitted signals that are caused by a device's unique hardware
``imperfections'' without \emph{any} knowledge about the nature of the
signal. 

We develop \system, a novel Deep Complex-valued Neural Network (DCN) that operates on \emph{raw} RF signals and is \emph{completely agnostic of the underlying applications and protocols}. We present two DCN variations: (i) Convolutional DCN (CDCN) for modeling full signals, and (ii) Recurrent DCN (RDCN) for modeling time series. Our system handles raw I/Q data from open air captures within a given spectrum window, without knowledge of the modulation scheme or even the carrier frequencies. This enables  the first exploration of device fingerprinting where the fingerprint of a device is extracted from open air captures in a completely automated way. 
We conduct an extensive experimental evaluation on large and diverse datasets (with WiFi and ADSB signals) collected by an external red team 
and investigate the effects of different environmental factors as well as neural network architectures and hyperparameters on our system's performance.
 \system is able to detect the device of interest in realistic settings with concurrent transmissions within the band of interest from multiple devices and multiple protocols.
While our experiments demonstrate the effectiveness of our system, 
especially under challenging conditions where other neural network architectures break down, we 
identify additional challenges in signal-based fingerprinting and provide
guidelines for future explorations. 
Our work lays the foundation for more research within this vast and
challenging space by establishing fundamental directions for using raw RF I/Q
data in novel complex-valued networks.

\end{abstract}

\section{Introduction}\label{sec:intro}
In recent years, the feasibility of identifying devices through browser or device
fingerprints has garnered significant %attention
attention~\cite{Eckersley:2010,laperdrix2016beauty,Acar:2013,starov2017xhound,Sjosten:2017,CaoLW17,MS12}.
Such techniques are not restricted to device characteristics and
specifications but can also fingerprint devices based on imperfections
inherent in the device's hardware~\cite{dey2014accelprint}.
In an alternative line of research, techniques have been proposed for fingerprinting devices based on unique
characteristics of the device's hardware that are exposed in their transmissions. A
remote network-based fingerprinting technique was presented in the seminal work by Kohno et
al.~\cite{kohno2005remote}, but could be prevented by software-level
defenses~\cite{brik2008wireless}. Brik et al.~\cite{brik2008wireless} proposed
a more robust but geographically-localized technique that focused on
transmitted radio frequency (RF) signals. However, their approach also relied
on protocol-specific information, and crucial practical factors were not
considered in the experimentation and evaluation. While both approaches pose
significant contributions, the diversity of wireless protocols in the wild
(which differ considerably across versions or
implementations~\cite{zeng1998glomosim}), coupled with the emergence of new
protocols and standards, hinders their applicability in many real-world
scenarios. Even the more generalizable techniques by Danev et 
al.~\cite{danev2012physical,danev2009transient} require, at least, some knowledge of higher-level but protocol-defined information, such as the frequencies of the carrier or sub-carrier signals or the bandwidth of the transmission channel. Inspired by the extensive prior research in the area, 
we explore novel methods to ameliorate current limitations and study 
the feasibility of fingerprinting with no prior knowledge of even 
the central carrier frequencies or modulation schemes involved.

Furthermore, while RF signals are inherently complex-valued, limitations of both 
theoretical frameworks and available development tools
have constrained RF-based research to real-valued conventional networks that severely limit
the representational power of generated models. However, recent breakthroughs
laid the theoretical foundations for complex-valued
networks~\cite{trabelsi2017deep} and paved the road for a new strain of deep
learning models which are able to work \emph{natively} with complex data. These
networks have higher representational capacity and thus more degrees of
freedom, allowing them to locate appropriate manifolds that are out of the
reach of conventional real-valued networks. 
In this work, we leverage the power of complex-valued deep learning
architectures, study their ability to operate on raw complex RF data, and
devise methodologies for training these novel networks.
Our models are trained on \emph{raw} I/Q data of signals captured under challenging environmental conditions in \emph{real} settings,
to show their applicability in practical and large scale deployments. In fact, \system is able to fingerprint devices using open air captures without any protocol preprocessing, and without knowledge of the underlying protocol or what carrier frequencies to look for. 

Specifically, our system guides the deep neural network towards identifying the artifacts in signals that are left by the unavoidable minor hardware variations or impairments (hereby referred to as ``imperfections'') that occur during the manufacturing process. 
To train our models and demonstrate  their performance under different
conditions, we utilize a plethora of datasets with as many as 1,000 devices.
With each sample corresponding to a captured transmission of only 64
$\mu$seconds, our total training set requires just \emph{6.4-51.2 milliseconds}
worth of captured signals per device. We show that when using only a 64 $\mu$s
long signal, \system can correctly fingerprint the device 72\% and 100\% of the
time for WiFi and ADS-B transmitters respectively. We also show that our
system is able to train a \emph{single} model for \emph{both} types and obtain
an accuracy of 82\% despite the drastic differences between the two signals
(i.e., protocols, transmission frequencies, and modulation schemes).

We find that \system is completely protocol-agnostic at the different 
spectrum ranges of our dataset and can handle devices that operate at 
multiple ranges (e.g., a device that transmits WiFi at 2.4GHz \emph{and} 5GHz)
without any prior knowledge, without relying on protocol implementation flaws like
predictable sequence numbers~\cite{vanhoef2016mac}, exposed device
identifiers~\cite{rupprecht-19-layer-two,musa2012tracking,DBLP:journals/corr/abs-1904-07623} or software-level data patterns (our dataset only includes control packets). 
Our system can be applied in a defensive capacity in a wide range of different scenarios,
including device-authentication for authorizing
access to resources (e.g., a cell tower or  access point), detecting
unauthorized or fraudulent devices in a restricted area, and unmasking spoofed
transmissions (including relay attacks or attacks using compromised certificates).
Nonetheless, our experimental evaluation is designed to abstract away the 
specificities of the deployment scenario and focuses on the central task of
identifying devices based on their transmitted RF signals.

Overall, our research presents new techniques for handling
raw RF I/Q data in their native complex format, highlights the challenges
of RF-based fingerprinting in \emph{real-world} settings,
and demonstrates novel directions for passively fingerprinting
devices without requiring any knowledge of the underlying protocols.
We believe that being able to fingerprint devices in an agnostic and scalable way can 
greatly augment current authentication schemes and defend against challenging 
attacks such as relay attacks or credential theft. 
Our evaluation explores numerous critical dimensions, yet
many interesting future research directions remain unexplored.
We hope that methodologies, network architectures, and experimental 
findings presented in this work will spearhead more research in the area.
In summary, our main research contributions are:

\begin{itemize}[leftmargin=*,nosep,noitemsep]
\item We present a novel device fingerprinting technique
that works on complex I/Q data representing \emph{raw} RF wireless
signals. We build a novel deep complex-valued network that is
\emph{protocol-agnostic}, designed to identify devices based on 
learnable characteristics from passively captured transmitted wireless signals.
This is the first, to our knowledge, system able to fingerprint devices using unprocessed raw signals in a range of frequencies of interest.

\item We detail a series of neural network architectures and configurations,
and explore their suitability for device fingerprinting. We derive
appropriate insights and highlight the inherent limitations of certain
architectures, which can help guide future research.

\item We provide an extensive experimental evaluation
using datasets from an externally organized red team.
Our experiments investigate multiple dimensions of signal-based device
fingerprinting and assess the impact of multiple factors under
different environmental settings and real-world conditions 
that previous work has not explored. We
demonstrate the effectiveness of our system's fingerprinting
capabilities under challenging and realistic scenarios,
using real-world, multi-device, open air captures of
RF signals.
\end{itemize}

\section{Background and Deployment Predicates}\label{sec:background}

\subsection{RF Primer}
We first introduce relevant background information
that facilitates comprehension of the following sections.
This includes basic telecommunication terms and notions
and technical details on the two signal families that we use throughout this paper.
As the methodologies involved in signal processing are sufficiently complicated, we refer readers without background in the area  to~\cite{proakis2007fundamentals} for a more complete overview.

\textbf{Baseband signals.} Signals are encountered in this form before being transmitted or after the receiver removes the carrier signal during the early stages of demodulation, centering the received signal around 0Hz.

\textbf{Modulation} is the process by which a transmitter encodes information so that it can be effectively and efficiently transmitted. The main idea is to encode the information from a \emph{baseband} signal by utilizing a \emph{carrier} signal of higher frequency. Information is encoded by modifying one or more of the carrier's characteristics: amplitude, phase, and frequency.

\textbf{I/Q data.} Signals are commonly represented in a format known as I/Q form~\cite{iq}. This is based on a methodology of representing a signal using a vector of complex numbers. The 'I' stands for the in-phase component whereas the 'Q' stands for the quadrature or out-of-phase component. Analytically it can be represented by the following equation:

\centerline{
$S(t)=A\cos{\theta t}+A\sin{\theta t} \rightarrow S(t)=I+Qj$
}

\noindent In essence any signal can be generated by a combination of in-phase and out-of-phase components and, hence, it is widely used in signal processing and telecommunications.

\textbf{SDR.} Software Defined Radios revolutionized the field of telecommunications as they allowed for the reception and transmission of RF signals with any modulation scheme and at any frequency while offering the convenience of defining everything through software despite utilizing the same underlying hardware. One of the key components behind this technology is the quadrature mixer.  The quadrature mixer can synthesize any signal using two signals with a phase difference of 90 degrees, and these quadrature signals can be conveniently expressed in analytical form as I/Q data. Figure~\ref{fig:quadrature_modulation} in Appendix~\ref{sec:app} depicts the process of transmuting I/Q data to a modulated signal using a quadrature modulator.

\begin{figure}[t]
\hspace{-0.51cm}
\includegraphics[width=1.1\columnwidth]{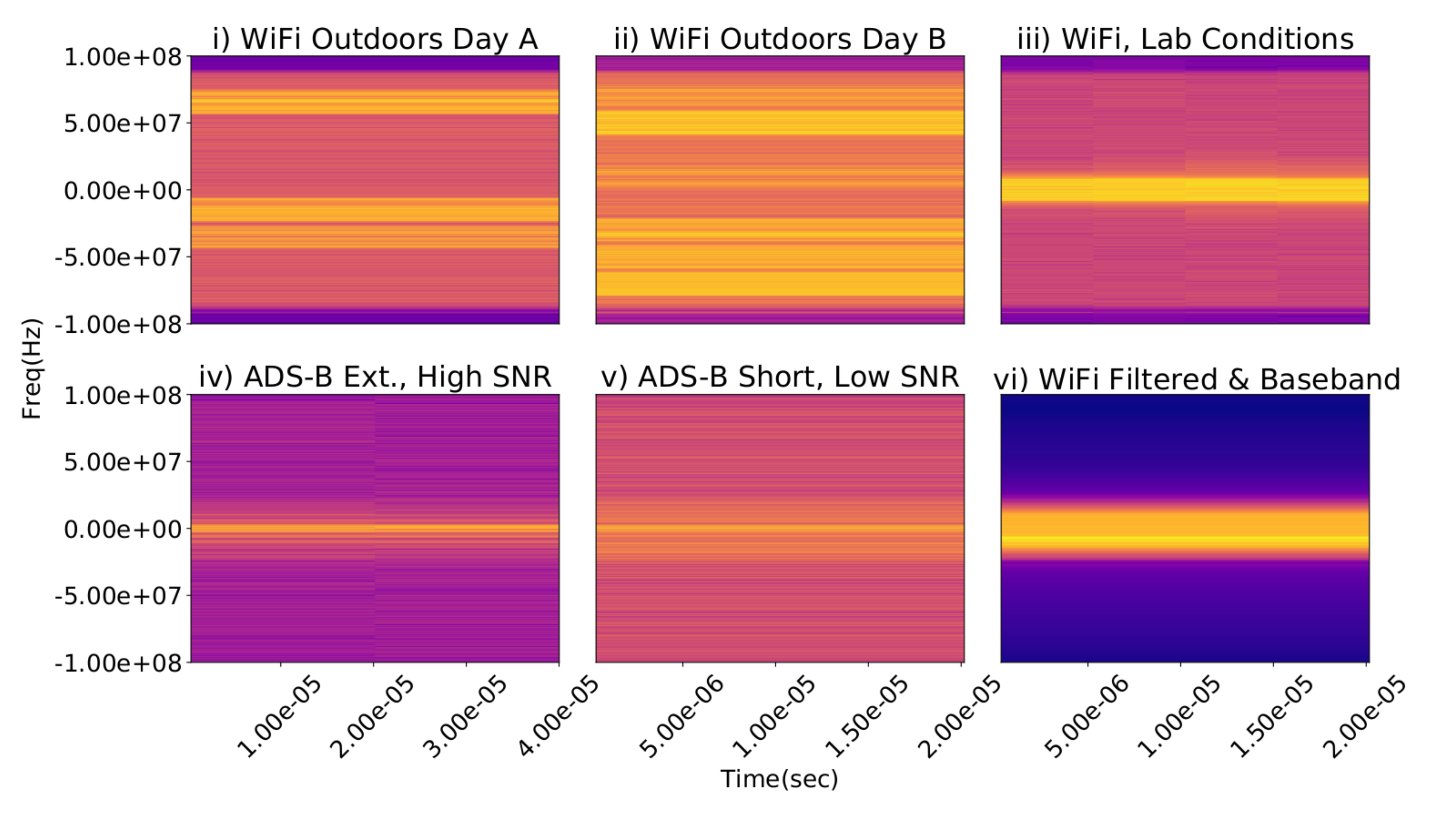}
\caption{Spectrograms depicting WiFi and ADS-B signals captured under
different channel environments and SNR. The outdoors WiFi signals depict the open air capture conditions.}
\label{fig:signal_overview}
\end{figure}

\textbf{Channel.} The medium through which the signal is transmitted, which can greatly alter signals in transmission and have a detrimental effect on them. A channel's characteristics are often ephemeral as they change due to a variety of environmental conditions, and the same signal transmitted through the same medium at different times can differ drastically. This inherent challenge is pertinent to our research goals, and we explore it in depth during our evaluation. 

\textbf{SNR.} Due to the transient nature of a channel's conditions, the Signal-to-Noise Ratio metric is useful in denoting the quality of the received signal by quantifying how much of a received signal is noise and what portion is actually useful.

\textbf{Spectrogram.} Captured transmissions can be visually represented as spectrograms. To obtain a spectrogram from raw I/Q data, a Short Term Fourier Transform (STFT) \cite{allen1977short} is performed over the entire captured trace. The SFTF outcome depicts the spectral content of the signal through time and one can easily identify drastic changes, the effect of noise, or different parts of the transmission. Figure~\ref{fig:signal_overview} shows a collection of spectrograms for both WiFi and ADS-B signals under different conditions to highlight the aforementioned effects. Plots (i) and (ii) show a WiFi signal captured from the \emph{same} device over two different days and (vi) depicts the signal after a basebanding and low pass filtering operation. Since the capture was conducted outdoors, the interference from other channels/devices is evident in the first two plots. In (iii) we show an unprocessed WiFi signal that was captured indoors where the SDR basebanded the signal without using an intermediate frequency (IF); one can notice that noise is still present since the signal has not been filtered. Plots (iv), (v) show both versions of ADS-B transmissions under different SNR (the x-axes are different for the two signals as the short version is almost half the transmission).

\textbf{WiFi.} WiFi signals are part of the 802.11 IEEE protocol family. While many variations exist, in practice the most frequently encountered follow the 802.11b and 802.11g specifications (which use  a central frequency of 2.4GhZ) and the newer 802.11n version (which utilizes the 5GhZ band). WiFi signals utilize the Orthogonal Frequency Division Multiplexing (OFDM) modulation which encodes information not in one carrier signal but in \emph{multiple} sub-carriers, where each sub-carrier can be modulated using amplitude, phase or frequency modulation schemes such as 16QAM, 16QPSK etc. 
The information content in a WiFi signal can vary significantly in size.
We direct the interested reader to~\cite{concepts_ofdm} for an excellent resource with more information.

\textbf{ADS-B.} A protocol of choice used by air crafts  and is scheduled to replace most RADAR systems according to the Federal Aviation Agency~\cite{faa_adsb}. It is encountered in two forms, either short or extended (64 or 120 $\mu$s). Apart from the header and the CRC, short messages only contain a unique aircraft identifier (24bits), whereas the extended version carries additional information on the altitude, position, heading,  horizontal and vertical velocity. ADS-B transmits at 1090MHz with 50KHz bandwidth or at 978MHz with 1.3MHz bandwidth. %The short and extended messages are also referred to as short and extended squitters.
The modulation employed in this protocol is referred to as Pulse Position Modulation (PPM), which encodes data by simply increasing or decreasing the width of the carrier pulses according to the value of the modulating signal at a given time. It is a much simpler modulation scheme than the one used in WiFi signals and the messages transmitted are either 5 or 10 bytes. An important characteristic of this protocol is that it is very robust to multipath effects and environmental noise. A more detailed overview and description can be found in~\cite{concepts_adsb}. While ADS-B packets contain a unique device identifier, certain transceivers have the option for anonymous mode where a randomized ID is sent~\cite{faa_adsb_faq}. Prior work has discussed how ADS-B data can  easily be spoofed to disrupt air travel safety~\cite{atanasov2013security,faragher2014spoofing}. As such, the ability to fingerprint and identify transmitters without relying on identifiers has significant operational value.

\subsection{Deep Learning Primer}
This section introduces the basic concepts required for following the deep learning networks and methodologies discussed in this paper. Readers familiar with the field may advance to the next section. An excellent detailed presentation of foundational concepts can be found in~\cite{Goodfellow-et-al-2016}.

\textbf{CNN.} Convolutional Neural Networks~\cite{lecun1998gradient} perform exceptionally well at detecting spatial points of interest and structural properties, and several variants have been used in a plethora of applications with great success, such as face and gender detection~\cite{ranjan2019hyperface}, semantic image segmentation~\cite{long2015fully}, speech recognition~\cite{zhang2017very} and stock prediction~\cite{zhang2019deeplob}. The basic idea behind a CNN is to perform a convolution operation between a weight vector of a given size (often called a kernel) and a subset of the input data, with the length of each convolution being the size of the kernel. Convolutional  layers are almost always used in conjunction with pooling layers to achieve dimensionality reduction without loss of information.

\textbf{RNN.} Recurrent Neural Networks are networks that excel at modeling  sequential or time series data. The core concepts behind them are the context and temporal relations between data points, potentially in different parts of the input. 
Two advanced forms of RNNs enhance the concept of context by introducing well-defined memory properties, namely Long Short Term Memory Models (LSTM) and Gated Recurrent Units (GRU). These advanced models allow the recurrent network to ``choose'' and ``memorize'' parts of the context that are significant for the model, emulating a form of memory. While the mathematical models behind each variant are different, both networks have been used in practice with significant success and similar capabilities~\cite{xingjian2015convolutional,graves2013speech,che2018recurrent,zhao2017lstm}. %The ability to selectively choose what to remember 

\textbf{Back Propagation Through Time.} As recurrent networks capture sequential properties and try to find temporal relations, the gradient descent step during the back propagation phase essentially attempts to link the current state with previous states in time. This is referred to as Back Propagation Through Time (BPTT)~\cite{werbos1988generalization}. RNNs face the problem of vanishing or exploding gradients for lengthy time sequences and while LSTMs are known to be more robust they can still suffer from the same problems albeit at much longer sequences~\cite{gers1999learning}.

\textbf{Complex-valued inputs.} Most networks currently used in practice operate on real-valued data. However, many problems are expressed in complex numbers, e.g., signals (RF or otherwise). Down-converting complex numbers to either the real or imaginary part limits the available information and introduces a miss-match between the model and problem at hand. To introduce complex-valued inputs and, thus, create complex-valued networks that theoretically have higher representational power (as they can express solutions in complex-valued manifolds) one needs to address many theoretical challenges and provide alternatives to elementary units such as activation units, or properly initializing complex-valued weights in hidden layers.

\textbf{Complex batch normalization.} Data normalization is commonly used for
efficiently training in most types of deep networks. However, when the
underlying problem lies in the complex domain, data requires special handling
as outlined in~\cite{trabelsi2017deep}. Briefly, one cannot perform two-way
independent normalization in the real and imaginary part of a complex number as
there is also information precisely in the relation of the real and imaginary
axes. This is highly evident if we consider the polar representation of  a
complex number $z = r\epsilon^{\iota\phi}$, where $\phi$ is the angle the
number forms in  the complex plane with the positive real axis, and is given by
$\phi=atan2(IM(z), Re(z))$. If one operates on either the real or imaginary
part independently, thus disregarding any correlation, we can introduce an
affine transformation which is not equivalent to normalization. To properly
handle the normalization process, we treat the complex numbers as 2D vectors
and the process as 2D whitening. We scale the \textbf{0} centered data by the
inverse square root of their covariance matrix \textbf{V} (the existence of the
inverse is guaranteed by the Tikhonov regularization; see
Appendix~\ref{theorem:tikonov}):
\centerline{$\Tilde{x} = (\textbf{V})^{-\frac{1}{2}} (\bold{x}- E[\bold{x}]),$}
\noindent where \textbf{V} is the 2x2 covariance matrix:
\begin{equation*}
    \textbf{V} = \begin{pmatrix} V_{rr} & V_{ri}\\
                               V_{ri} & V_{ii}    
                \end{pmatrix} = \begin{pmatrix} Cov(\Re{(x)}, \Re{(x)}) & Cov(\Re{(x)}, \Im{(x)})\\
                                                Cov(\Re{(x)}, \Im{(x)}) & Cov(\Re{(x)}, \Re{(x)})\\
                                \end{pmatrix} 
\end{equation*}

The complex batch normalization layer is defined as:
\begin{equation*}
    BN(\Tilde{\textbf{x}}) = \gamma \Tilde{\bold{x}} + \beta, \, where\, 
    \gamma = \begin{pmatrix} \gamma_{rr} & \gamma{ri} \\
                             \gamma_{ri} & \gamma_{ii}
             \end{pmatrix}
\end{equation*}

\textbf{Complex activation units.}
As in the normalization case, activation functions must also be adapted to handle complex-valued numbers. Several activation units have been proposed such as ModRelu~\cite{arjovsky2016unitary}, $\mathbb{C}$ReLU~\cite{trabelsi2017deep} and  zReLU~\cite{guberman2016complex}. These functions preserve differentiability, at least in part, to maintain compatibility with back-propagation. As shown in~\cite{hirose2012generalization}, full complex differentiable functions are not strictly required. While many activation functions have been proposed, we evaluate the following complex activation functions:
\begin{align*}
  %  modReLU &= ReLU(|z|+b)\epsilon^{\iota \theta_z} \\
    \mathbb{C}ReLU   &= ReLU(\Re{(z)}) + \iota ReLU(\Im{(z)}) \\
    zReLU   &=  \biggl\{ \begin{array}{ll}
                        z & if \theta_z \in [0, \pi/2],  \\
                        0 & elsewhere\end{array}
\end{align*}

\textbf{Complex weight initialization.}
Appropriate weight initialization is often considered helpful for the training process of networks. It can help avoid the risk of vanishing or exploding gradients and facilitate a faster convergence. In the complex case, as derived in~\cite{trabelsi2017deep}, the variance of the weight vectors is given by Var(W) = $2 \sigma^2$, where $\sigma$ is Rayleigh distribution's only parameter; $\sigma$ can be selected according to any initialization criterion such as those described  in~\cite{glorot2010understanding,he2015delving}. Note that as the variance is not affected by phase, one can initialize the phase $\phi$ of W as $\phi_W \sim U[-\pi, \pi]$.

\subsection{Deployment Predicates}

The techniques presented in this work can be applied to a wide range of scenarios. %(we mention some examples in Section~\ref{sec:future}).
Nonetheless, for ease of presentation we will assume that role of an entity
aiming to identify the sources of wireless signals (be it electronic devices or
air crafts). We assume that this entity passively collects
wireless signals within a certain area, without interacting with or actively
probing the devices. Similarly we do not require the user to interact with a
specific resource. The only assumption is that the user's device is
transmitting some form of wireless signal; in our experimental evaluation we
use WiFi and ADS-B traces, but this could be applied to other types as well
(e.g., Bluetooth, GSM, 4G, etc ). 
Depending on the scenario, the application can include the deployment of
multiple antennas within a larger physical area
(e.g., for authenticating devices connecting to an organization's restricted wireless network)
or may focus on a specific small
space (e.g., for detecting rogue access points using stolen certificates).

\section{System Design and Implementation}\label{sec:methodology}
Here we present an overview of \system, and discuss the motivation behind our design and implementation choices.

\textbf{Design constraints.} The main goal of \system is to operate on raw I/Q data so as to decouple the fingerprinting capabilities of our system from protocol or software specificities (which can differ across versions~\cite{zeng1998glomosim}) and implementation flaws. It also removes the burden of manual analysis that would otherwise be required if new or custom protocols were encountered.
To that end we limit all data preprocessing to general RF methodologies (such as applying low or band pass filtering) and data augmentation techniques that can be applied to any RF signal regardless of protocol (such as decimation).
\begin{figure*}[t]
\centering
\subfloat[Convolutional Deep Complex Network (CDCN) architecture.~\label{fig:dcn_rfnet}]{\includegraphics[width=0.85\textwidth]{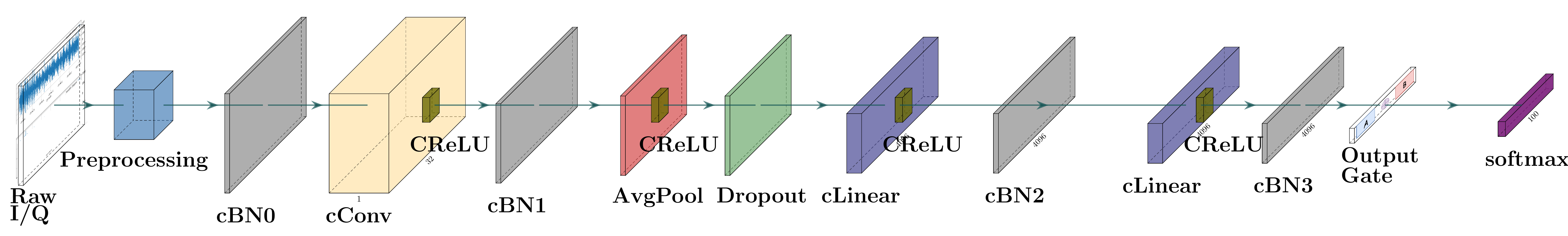}
}
\vfill
\subfloat[Recurrent Deep Complex Network (RDCN) architecture.~\label{fig:lstm_rfnet}]{
\centering
\includegraphics[width=0.85\textwidth]{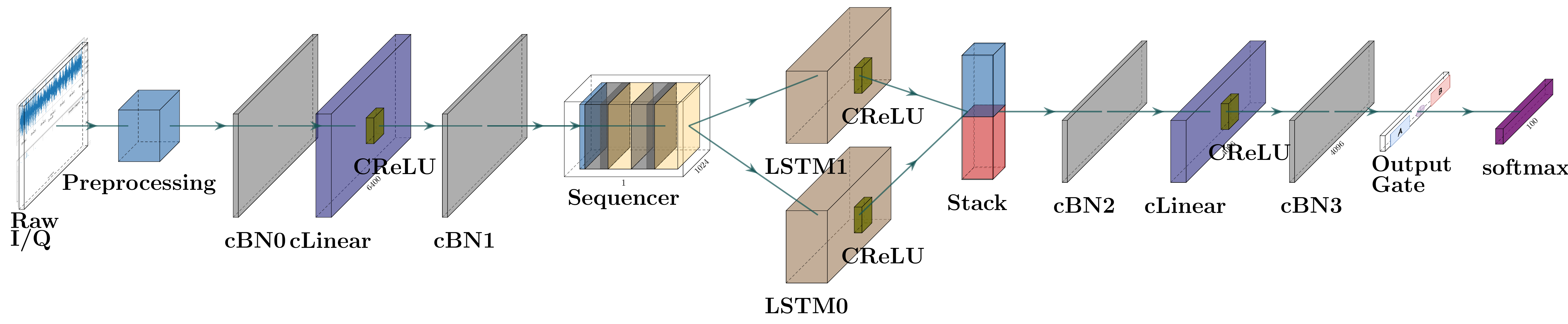}
}
\caption{Convolutional Deep Complex Network (CDCN): the Complex Convolutional Layer combines both I/Q pairs in two mixed channels \textbf{(top)}. The Recurrent Deep Complex Network (RDCN) is based on the powerful LSTM variant; while initially the I/Q pairs are in two different channels the Complex Linear layer binds them together, and its output contains two mixed channels \textbf{(bottom)}. The schematic was produced with PlotNeuralNet~\cite{plotnet}.}
\label{fig:rfnet_both}
\end{figure*}

\textbf{CDCN.} We first develop a complex-valued DCN network (Figure~\ref{fig:dcn_rfnet}) similar to the shallow MusicNet from the recent work of Trabelsi et al.~\cite{trabelsi2017deep}, which was successfully applied to acoustic data. We draw inspiration from this network because even though acoustic and RF signals are different in nature (one is mechanical and the other electromagnetic) they share many similarities. Both can be modeled as a 1D complex vector of samples, and many of the techniques, methodologies and signal processing theories are applicable to both in practice (low/bandpass filtering, decimation, etc). This also poses as a comparison baseline for our novel LSTM-based approach.

\textbf{RDCN.} Since RF signals can be expressed as a time series and have temporal coherency, we hypothesize that networks that capture such temporal features will perform well at analyzing RF data. Consider, for instance, the case of amplitude modulation where voltage changes from a higher value to a lower value must pass through intermediate values (albeit ephemerally). The precise  nature of this transition is based on the device's electronics and this is precisely what we aim to capture.  Our second complex network elevates LSTM layers, and employs a custom build layer we call the ``Sequencer'', as shown in Figure~\ref{fig:lstm_rfnet}. A known problem for RNNs is the BPTT problem; as RF signals can expand to extremely large time series ranging to several thousands of samples, simply operating on a point to point basis is infeasible. While LSTM's memory mechanisms help alleviate  the problem by providing alternative ways for the gradient to flow back through time, they can still break down under sufficiently large signals~\cite{gers1999learning}, as the memory cells grow to crippling proportions. Drawing upon the conclusions of Gers et al.~\cite{gers1999learning} to combat this problem we have designed the Sequencer layer, which splits the signal -that is modeled as a 1D time sequence of I/Q pairs into a number of smaller vectors. The generated vectors are then served as inputs sequentially to the LSTM in a proper temporal order as shown in Figure~\ref{fig:decimation-sequencer}, allowing the LSTM to ``see'' the same amount of data but over less timesteps. This transmutation of a longer vector into a collection of smaller vectors changes the nature of the signal by transforming it into different dimensions. As such, the size and number of the newly generated vectors are critical as they might not preserve the qualities of the original signal.

 % \includegraphics[width=\columnwidth]{images/dcn.pdf}
 % \caption{Deep Complex Convolutional Network Architecture for RFNET}
 % \label{fig:dcn_rfnet}
\subsection{Real, Imaginary and Mixed Data Channels}
Due to limitations of the underlying frameworks, the fundamental data-holding structure (i.e., \emph{tensors}) cannot contain complex numbers. To overcome this our tensors contain the real (``I'') component in one dimension and the imaginary (``Q'') component in another, akin to~\cite{trabelsi2017deep}. Initially, our input channels are clearly separated between real and imaginary components containing real-valued float numbers. This separation ceases once the data is pushed through the first layer. The linear, convolutional layers operate on both channels utilizing the equations described previously, and the output channels exiting these layers contain two channels with data containing information from both the real and imaginary parts.

These output channels no longer solely represent the real and imaginary parts of the number, but are instead mixed information-wise. To differentiate from the prior ``pure'' input channels we refer to the new channels as ``A'' and ``B''. To illustrate the qualitative difference between these sets of channels, consider the output of the complex linear layer:
\begin{equation}
    L_W^{\mathbb{C}}(\textbf{h}) = 
    \begin{bmatrix} \Re{(\textbf{W} \cdot \textbf{h})}\\
                    \Im{(\textbf{W} \cdot \textbf{h})})
    \end{bmatrix} = \begin{bmatrix} 
                        \textbf{A} & \textbf{-B}\\
                        \textbf{B} & \textbf{A} 
                    \end{bmatrix} \cdot \begin{bmatrix} 
                                        \textbf{x}\\
                                        \textbf{y}
                                    \end{bmatrix}
\end{equation}
\vspace{-1mm}
The outcome of this equation contains values that are qualitatively mixed and stores them as: 
\begin{itemize}
	\item channel \textbf{\emph{A}}$=R*R -I*I$ 
	\item channel \textbf{\emph{B}}= $R*I +I*R$
\end{itemize}

Mathematically, the outcome of this operation is still real-valued in one channel and imaginary in the other. However, information-wise the channels contain mixed data since the imaginary values are used in tandem with the real valued parts.

\textbf{Output channels.}
The output of the network is contained in two mixed channels \textbf{\emph{A}} and \textbf{\emph{B}}. Trabelsi et al.~\cite{trabelsi2017deep} utilize the output of channel \textbf{\emph{A}}, and discard the output of channel \textbf{\emph{B}} in the shallow version of their proposed network but utilize both channels through a linear dense output layer in the deep version of their model. Since we are working with novel RF data we explore three possible configurations of the outputs so as to identify the optimal way for leveraging the mixed output channels: (i) only the output of channel A (per prior work), (ii) we introduce parameters $G_a$ and $G_b$ and sum the channels, allowing the network to modify the parameters and (iii) we employ a convolutional layer for joining the two channels into one output channel of identical dimensions.
It is important to note that while Wolter et al.~\cite{wolter2018gated} follow a similar scheme for the gates that control the memory of the Complex GRU that should not be confused with our parameters which act as a weighting factor for the two output channels. Our results show that both channels are actually used in practice and the network keeps both parameters in close proximity.

\subsection{Layers}
\textbf{Complex layers.} We use the complex batch normalization, complex convolutional, and complex linear layers proposed by Trabelsi et al.\cite{trabelsi2017deep}. The input of the initial convolutional and linear layers is split in two channels  that contain the (purely) imaginary and real part in each channel. However, as aforementioned, the output of these layers contains  mixed information from both imaginary and real parts of the signal.

\textbf{Activation units.} To identify the optimal activation unit we evaluate both ZReLU and CReLu. After the initial complex convolutional/linear layer the output channels contain mixed information. We found that ZReLU, which activates on the first quadrant of the Cartesian plane, is not optimal for the mixed nature of the data and CReLU provides better results.

\textbf{LSTM.} Our LSTM layers are unmodified real-valued LSTMs, but the input they receive is mixed. We utilize two LSTMs in parallel and provide the \textbf{\emph{A}}, \textbf{\emph{B}} channels as input respectively. The output used for the classification is the context from each LSTM during the \emph{final} step. The internal context is preserved for the entire epoch and reinitialized to a clean context at the start of each epoch. The computational graph is preserved within each minibatch but detached -yet the context \emph{values} are still preserved across minibatches so as to avoid backpropagation across minibatches.
During our experiments we found that the RDCN architecture was significantly harder 
to train, corroborating the findings of \cite{pascanu2013difficulty}.

\textbf{Preprocessing step.} Our data preprocessing is minimal and comprises of: (i) basebanding the signals \cite{proakis2007fundamentals}, and (ii) applying a $3^{rd}$ order Butterworth bandpass filter~\cite{butterworth1930theory} with cutoffs at 25KHz and 20MHz. Our insight behind the lower order Butterworth filter is to allow a smooth cutoff and include potential discriminating effects emanating from the imperfections inherent to the transmitter's filters. The precise lower value was calibrated experimentally. No other data preprocessing or other protocol-specific treatment takes place.

\subsection{Data Augmentation}
Existing methodologies for processing image or audio data contain a multitude of data manipulation techniques for creating new synthetic instances by increasing robustness and reducing generalization errors. For instance, a well established approach for creating synthetic image training data is to rotate the image, scale it or add/remove noise. Prior work has shown that such practices improve the network~\cite{ding2016convolutional,lv2017data,salehinejad2018image}. We adapt common data augmentation practices and evaluate their applicability and limitations in the realm of RF data.

\begin{figure}[t]
\cheatspace
\centering
\includegraphics[width=0.8\columnwidth]{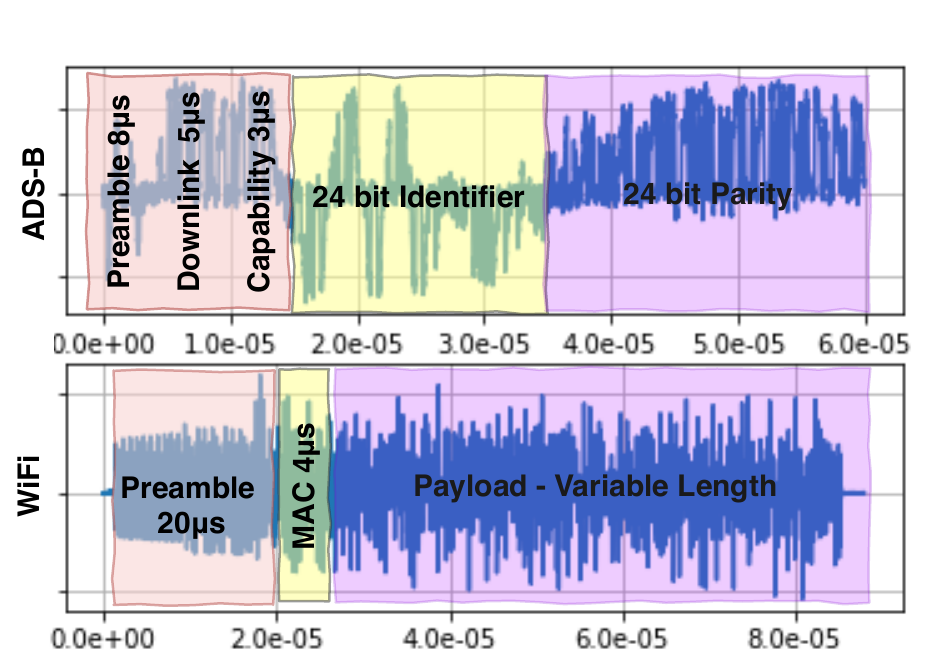}
\caption{Signal breakdown and unique identifier position for the ``I'' component over time.}
\label{fig:signal_identifiers}
\end{figure}

\begin{figure}[t]

\centering
\includegraphics[width=\columnwidth]{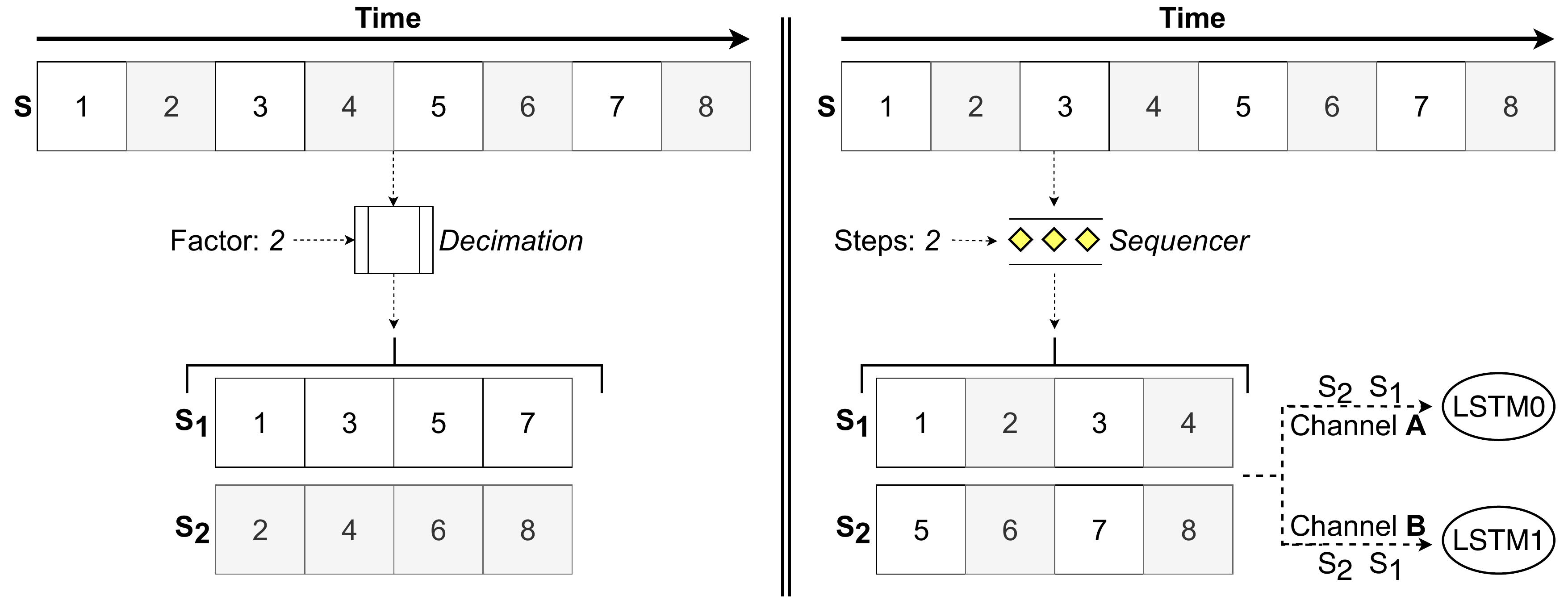}
\caption{Decimation with a factor of 2 generates two new samples of half the
original length \textbf{(left)}. The Sequencer converts the signal into a
series of vectors (here a vector of 2 vectors) that are sequentially
entered in the LSTM in proper temporal order, i.e, the samples will be seen
in the \emph{same} order as if the original signal was provided unprocessed
\textbf{(right)}.}
\label{fig:decimation-sequencer}
\end{figure}

\textbf{Decimation.} A widely used methodology in signal processing is to
reduce the computational load by keeping every $M_{th}$ sample of a signal
where $M$ is called the decimation factor. In order to avoid aliasing a low
pass filter must be applied before the decimation operation. For neural
networks apart from reducing the computational load, decimation also reduces
the number of network parameters since the input signal is reduced according to
the decimation factor. In our work decimation has the additional role of
augmenting the dataset since the decimated signals are used as extra samples.
The decimation factor must be an integer value and obey the lower limit of the
Nyquist theorem~\cite{nyquist1928certain} to be able to correctly reconstruct
the signal. Decimation can be seen as reducing the rate of sampling by a ratio
equal to the decimation factor $M$, thus the maximum ratio must be the number
that allows a sampling rate with at least $F_s\geq2F_{max}$, where $F_{max}$ is
the maximum frequency required. For example in the case of WiFi where the
bandwidth is 20MHz the minimum required sample rate is 40MSPS (Mega Samples Per
Second); since our sample rate is 100MSPS per channel and the factor $M$ must
be an integer the only allowed value is two. Increasing the decimation factor
beyond that limit reduces the highest frequency one may reconstruct, leading to
potentially missed information. To establish whether this constrain also
applies to the problem at hand (which is to fingerprint the transmitter rather
than faithfully reconstruct the signal) we add this parameter to our
hyperparameter exploration. Figure~\ref{fig:decimation-sequencer} (left) shows
an illustration of decimating a signal composed of 8 data samples with a factor
of two. We also adapted other common data augmentation techniques 
but did not find them to be useful in this context. We
provide more info in Appendix~\ref{app:data_augmentation}.

\textbf{Randomized Cropping.}
Our main design goal is for \system to be protocol-agnostic and to not learn software-level characteristics. To guide the network away from latching onto protocol identifiers, we crop each sample in every minibatch into N parts and randomly choose, without replacement, one part to train on.
This approach forces the network towards selecting features that are common in different parts of the signal since it has to minimize the error for every part it encounters during training. The same procedure is followed for testing as well, and each test signal is fragmented into N parts and a random piece is chosen for classification.

Both WiFi and ADS-B signals used in our evaluation, contain identifiers that exist in predetermined positions and have different sizes (see Figure~\ref{fig:signal_identifiers} in Appendix~\ref{sec:app}). In our current experiments we find that $N=6$ provides a good balance between having a sufficient part of the signal and making sure that the majority of the generated random crops do not contain any unique identifier. Our technique is protocol agnostic as we do not make assumptions on the precise location of the unique identifier in the signal, choosing a random part when training or testing. In other words, during our experiments 83.3\% of each testing and training signal is randomly discarded, which prevents the system from latching on to identifiers without us revealing information to the system about the position of the identifier within a specific type of packet (as that would implicitly teach the system information about each protocol).  
By following this scheme, the network is forced to identify the device without learning superficial features that correspond to identifiers, and it does so regardless of where such an identifier is located in the signal. 
The results solidify our hypothesis that our approach captures the essential characteristics of each device; if the networks had learned to predominantly discern a signal through its identifier, the accuracy results reported with random signal cropping would arguably be significantly lower than those presented in this work.

\section{Dataset Description}
\label{app:dataset}

Here we provide details of the datasets used throughout our experiments, which
are comprised of signal traces captured under a variety of conditions over a
period of 13 days. This was part of our external evaluation by a red team (they were 
in charge of creating the dataset -- performers were not part of IRB-related processes).
Data is stored in the SigMF format, which we describe in Appendix~\ref{sub:sigmfdata}.

The RF signals were captured from 96 different Tektronix RSA5106B each equipped
with a HG2458-08LP antenna, and four USRP B210 SDRs. Signals were captured at
100MSPS\footnote{The overall sampling rate is 200MSPS split between both  ``I''
and ``Q'' components leading to 100 MSPS per channel.} which is well beyond the
minimum sample rate required to reconstruct even the most demanding signal
category in our dataset, i.e., WiFi signals. Finally, the captures were
conducted with both vertical (90\%) and horizontal (10\%) antenna polarization,
5\% of the signals are captured in both forms. The captured signals have
variable lengths according to the protocol and the message type. ADS-B signals
have a length of 6,400 (64$\mu$s short) or 12,000 (120$\mu$s
extended) whereas for WiFi length is extremely variable (as per specification),
ranging from a few thousand I/Q pairs to 272K I/Q pairs in our data. As our
goal is to create a system without any knowledge of the underlying protocols,
we perform our experiments with vectors of 6,400 samples (I/Q pairs) for both
WiFi and ADS-B. We do not differentiate between protocols by adding
more data points for WiFi signals, as that would increase the informational
content the system sees for one type and could guide it to separate signals due to superficial characteristics (e.g., different signal lengths). At our sample rate 6,400 complex data points correspond to to 64$\mu$s of signal capture. 

\textbf{WiFi-1.} This dataset spans all 13 days and data was captured in both indoor and outdoor settings.
It includes 4TB of signals that have been collected from 53,853 unique devices from a variety of manufactures. The majority of devices are Apple iPhones ($\sim$30\%) and the rest are from different manufacturers and types (smartphones, laptops, tablets, drones). The signals captured are mainly from 2.4Ghz  but a significant portion ($\geq20\%$) is also from 5Ghz emitting devices; some devices transmit in both bands. 
Device labeling was done using the MAC addresses, as MAC address randomization was disabled to facilitate the labelling process of such a massive and diverse set of devices. 
The real-world capturing process has also resulted in additional interference and ``noise'' in the captured signals from other protocols that transmit at those ranges (e.g., Bluetooth), which poses an additional challenge for protocol-agnostic fingerprinting. Furthermore, the open air capturing was designed to eliminate or minimize environmental sources of bias that could assist the fingerprinting systems. Specifically, while the deep learning models could potentially learn environmental factors or location-specific artifacts, this was counteracted by: (i) the weather and other environmental factors (e.g., physical obstacles that alter signals) changing over the course of 13 days, (ii) using devices that are not stationary (as those commonly used in prior studies) but mobile devices during normal operation where they change multiple locations per day while transmitting their signals, and (iii) the antennas being deployed in different locations which also changed through time.
As such, this dataset allows for the evaluation of our proposed models across different capturing conditions and for large-scale deployments where many devices need to be tracked. 

\textbf{WiFi-2.} This dataset is formed from a set of 19 identical devices with
the same firmware version installed. The signals were collected in a lab
(indoors) and the devices were set up to transmit the same data and configured
with the same MAC address. The dataset includes $\sim$200K signals and requires
17GB of storage. %evenly distributed amongst all the devices.  Here the
Labeling was done manually: traffic was captured from one device and
labelled, then the next device was used and labelled, etc. It is important to
note that while all devices have the same MAC address, and devices transmitted
the same data, ephemeral identifiers in the traffic will differ (e.g.,
timestamps). The main purpose of this dataset is to demonstrate that \system
does not learn specific identifiers and is unaffected by other software or
content-related artifacts, but is instead capable of identifying the artifacts
left by a device's unique hardware imperfections in a transmitted signal.

\textbf{ADSB-1.}
This dataset was collected over a span of 10 days and contains approximately an equal amount of short and extended signals. The transmitters were airplanes and we have collected signals from 10,404 unique sources. The captured data contains a plethora of different airplane positions, velocities, altitude and SNR values. Labeling was done using the unique 24bit identifier encoded in each message. The entire dataset includes 3.5M signals and is approximately 7TB. The outdoors collection took place at a single location.

\textbf{Realistic environmental conditions.} Numerous  factors affect the quality of a recorded signal; channel effects (e.g., multipath, hysterisis), environmental effects (e.g., outdoors vs indoors), the type of transmitter (e.g., cellphones, UAV), capture conditions (e.g., antenna polarization), will vary in reality. Our dataset was generated so as to contain signals representing \emph{all} of the above conditions. We believe that this plethora of different capture parameters is a critical requirement for evaluating a system's ability to fingerprint devices \emph{in practice}, as ideal indoor-lab settings alone are not indicative of actual deployment settings. Furthermore, the signal captures can contain multiple concurrent transmissions from different devices, thus, introducing additional noise. Another important feature is that our training and testing data are not from a contiguous time window (apart from WiFi-2), further increasing the realism of our testing scenarios.
Prior studies have evaluated their proposed systems using datasets that reflected some but not all of the aforementioned parameters. To our knowledge, we are the first to present a study that experimentally evaluates a system using a dataset that contains signals captured in such diverse conditions \emph{and} quantifies their effect.

\section{Experimental Evaluation}\label{sec:evaluation}

Here we provide an extensive evaluation that explores
\system's performance under different environmental factors and
across multiple dimensions of our model's design and implementation. The
combinations of factors and the specific details (e.g., number of training and
testing samples) in each experiment were dictated by a red team evaluation
process organized externally (we have provided more information to the PC chairs).
All the experiments presented, unless otherwise noted, were
performed on captured signals with a duration of 64$\mu$s represented by 1D
vectors of 6,400 complex numbers. In all experiments the testing and training
data are from different transmissions. The reported accuracy in each case denotes 
the percentage of \emph{target} devices that were identified correctly, thus,
presenting a lower bound of the actually accuracy of our models. For instance, 
consider an example where the test signal
capture for device A also contains background transmissions from a device B
which is one of the other devices used in the experiment. If our system
labels the test signal as ``Device B'' we will count that as a misclassification,
despite Device B being one of the classes and present in
the capture.

\textbf{Hardware specifications.} Experiments were performed on a
Supermicro SuperServer 1029GQ-TVRT equipped with 2 Intel Xeon Bronze 3104 (6
cores at 1.7GHz each) with 96GB of DDR4 RAM and two Nvidia Tesla V100 SXM2 GPUs
with 16GB of VRAM each. All of the architectures 
occupied a maximum of 8GB on the GPU; the multiple units enabled the concurrent
training of different models and reduced the time required for the
hyperparameter exploration.

\textbf{Models.} We present two real-valued conventional networks, namely ANN
and CNN, and two complex-valued networks, namely CDCN and RDCN shown in
Figure~\ref{fig:rfnet_both}. We use the real-valued networks as baseline for
evaluating the potential of the complex-valued variants and drawing a baseline
of comparison using all the information present in a signal (properly bound
using the covariance matrices, and the usage of complex weights and complex
activation units). 

\textbf{Machine learning configuration.}
The networks were trained over 100 epochs with a batch size of 32. We found
that adaptive learning rate annealing is helpful and, thus, reduce the learning
rate by $1e^{-1}$ after every 10 epochs where the network fails to improve its
validation accuracy. We also allow \emph{early stopping} once the learning rate
reaches $1e^{-7}$. The results we report in this section represent the best
generalization a ccuracy over 100 epochs or if the early stopping criterion is
met. This falls in line with evaluation approaches in the deep
learning literature when handling large and complicated neural networks
(e.g.,\cite{prechelt1998early}, \cite{bishop2006pattern},
\cite{goodfellow2016deep}, \cite{zhang2016understanding}). This is standard
practice for obtaining unbiased error estimations when facing significant
computational requirements for training such networks and exploring different
architectures, configurations, and datasets, as opposed to k-fold  validation
approaches that are common for other machine learning techniques. We evaluated
both Cross Entropy and Binary Cross Entropy and found that the  later improves
the convergence of the model and produces better results.
We evaluated both the Adam optimizer and SGD in a subset of the experiments and
found that Adam with amsgrad disabled is superior.  Real-valued layers are
initialized using uniform distribution based on the work by Glorot et
al.~\cite{glorot2010understanding}, while complex-valued layers leverage
complex weight initialization. We performed no normalization and
standardization of the data during preprocessing, and relied on the first Batch
Normalization layer that performs a similar computation. The hyperparameters
that we explored include several variables such as the optimal decimation
factor, the optimal sequencer time step and the number of hidden units in the
RDCN layer. The hyperparameter space was explored using
Spearmint~\cite{snoek2012practical}, and the optimal values obtained
corresponded to the best accuracy using a validation dataset comprised of both
WiFi and ADS-B devices. Our methodology and details on the obtained
hyperparameters can be found in Appendix~\ref{app:hyperparameters}.

\textbf{Hyperparameter exploration takeaways.}
While our experiments show that the hyperparameters greatly affect the
learning capabilities and accuracy of each architecture, our
exploration reveals that the RDCN network is the most ``sensitive'' to
hyperparameters.
Its performance can range from not learning anything (i.e., getting results
equivalent to randomly guessing the source of the signal) to obtaining highly
accurate results. For RDCN we find that the most crucial
parameter, besides the learning rate, is the number of time steps produced by
the sequencer; this indicates that restructuring the 1D time series vector into
a tensor of vectors directly influences the representation of the signal,
effectively changing an N-length one dimensional signal to an N/M vector length
of M-dimensional vectors. The best results were obtained for a length that
corresponded to approximately 1$\mu$s worth of data, given our sampling rate
that corresponds to 100 values per timestep (or 50 when a factor 2 decimation
is used).

\begin{table*}[t]
\center
\caption{Accuracy in regards to temporal relation of the collected signals, using the WiFi-1 dataset.}
\begin{tabular}{lcccccccccc}
\toprule
\textbf{T-train} & \textbf{T-test} & \textbf{Environment} & \textbf{Classes} & \textbf{Samples (Train)} & 
\textbf{Samples (Test)} & \textbf{ANN} & \textbf{CNN} & \textbf{CDCN} & \textbf{RDCN} & \textbf{RDCN (top5)}\\
\toprule
\rowcolor{Gray}
Day-1 & Day-1 & Indoors & 100 & 800 & 200 & 39\% & 66\% & 74\% & 72\% & 90\% \\ %experiment 3C -- FCC column for all
Day-1 & Day-1 & Outdoors & 100 & 800 & 200 & 38\% & 52\% & 66\% & 63\% & 89\% \\ %experiment 3D
\rowcolor{Gray}
Day-1& Day-1   & Both & 100   & 800 & 200 & 32\% & 61\% &  71\% & 70\% & 91\% \\ %experiment 3B
Day-1 & Day-2 & Both & 50  & 100 & 25 & 2\%  & 3\% & 7\% & 10\% & 30\% \\ %experiment 3A
\rowcolor{Gray}
Day-1,2 & Day-3 & Both & 65 & 218 & 54 & 4\% & 10\% & 12\% & 21\% & 44\% \\ %experiment 3E
\bottomrule
\end{tabular}
\label{tab:time-results}
\end{table*}

\subsection{Signal Preprocessing}\label{subsec:sigproc}

\begin{figure}[t]
\centering
\includegraphics[width=\columnwidth]{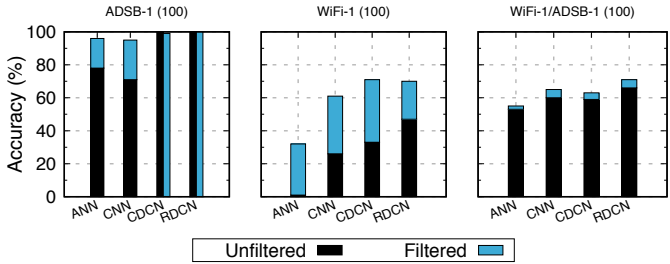}
    \caption{Preprocessing effect on accuracy, for 100 devices. While all networks benefit from preprocessing, CDCN and RDCN are more robust to learning from unprocessed RF data.}
\label{fig:filtering}
\end{figure}

First we explore how preprocessing the signal prior to feeding it to \system affects accuracy.
To that end we compare the accuracy obtained when the different neural architectures are fed completely
unfiltered data as opposed to data that has undergone generic and standard telecommunications preprocessing,
namely baseband and passband filtering.
Figure~\ref{fig:filtering} shows that filtering 
and basebanding are crucial for the WiFi signals. Apart from higher susceptibility 
to outdoor conditions and channel effects, the inherent ability of WiFi 
transmitters to broadcast in different communication channels mandates the use of 
basebanding. Interestingly, we find that the complex-valued networks are able to 
``see through'' the clutter induced by concurrent WiFi transmissions and ambient noise,
locating the intended signal even in the raw pre-processed signal case where the signal 
is not even basebanded and the captured instance between training and testing could 
be in \emph{different} central frequencies.
It is evident however that performing at least 
the basic signal processing can
greatly improve the accuracy, especially in the weaker real-valued networks.

\subsection{Multi-protocol networks} %THIS IS NOT RS6 approach

\begin{figure}[t]
\centering
\includegraphics[width=\columnwidth]{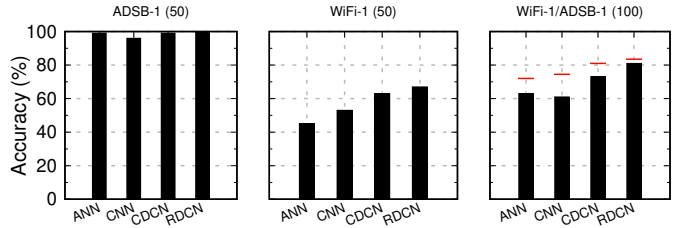}
\caption{Accuracy comparison of two stand-alone single-protocol networks versus a protocol-agnostic 
approach for all signal sources. The red lines denote the ``ideal'' accuracy, which is the average
of the two stand-alone networks.}
\label{fig:multi-protocol}
\end{figure}

Apart from the positive effect of filtering, Figure~\ref{fig:filtering} also indicates the challenge
of building a network for signals from multiple protocols.
When the network is trained on a mixture of two signal types, performance is lower
than the average accuracy of the two stand-alone networks (which one might hope to achieve).
To further explore this effect we consider the following experiment. We assume a scenario 
where 50 WiFi and 50 ADS-B sources exist, and want to quantify the performance impact of 
a protocol-agnostic approach where no explicit information is provided to the network for differentiating 
between them. Specifically, we quantify the accuracy of two stand-alone 
versions of \system, each trained and tested on 50 classes of a single protocol,
and compare it to a protocol-agnostic version that has to handle all 100 classes.
As can be seen in Figure~\ref{fig:multi-protocol}, the multi-protocol network is ``weighed down''
by the WiFi protocol and is not able to reach the average accuracy of the two standalone networks.
It is important to note that while the increase in the number of classes from 50 to 100 can affect
the performance, the presence of multiple types of protocols has a dominant impact (as in Figure~\ref{fig:filtering}).
We find that the RDCN architecture not only achieves the highest accuracy, but also
presents the smallest accuracy loss for the multi-protocol system, which further highlights its
effectiveness and suitability for challenging settings where multiple types of signals are emitted
from sources.

\subsection{Channel effect: spatial implications}

Due to the transient nature of the channel through which a signal is being transmitted,
we explore how \system's accuracy changes based on the spatial conditions of the 
different signals. We break down our experiments on the WiFi-1 dataset in the first two rows of
Table~\ref{tab:time-results}. We start by using indoor data that was captured
on the same day, for 100 different devices. % and using a 4:1 ratio of testing to training data. 
Our RDCN network is able to correctly identify the device in 72\% of the cases,
while the correct device is in the top five for 92\% of the tests. When we replicate this
experiment using signals that were captured outdoors, accuracy drops by 4-9\%, due to the 
impact environmental factors have on the transmitted signals. When training and testing 
using a mixture of indoor/outdoor signals \system performance is roughly in the middle 
between the indoor and outdoor settings. Once again, RDCN vastly outperforms real-valued
networks, demonstrating our system's robustness against environmental factors -- if our model relied on location-specific artifacts for identifying devices its performance would have significantly deteriorated in this experiment.

\subsection{Channel effect: temporal implications}

Prevalent ambient environmental differences greatly affect the morphology of each signal
as shown in Figure~\ref{fig:signal_overview}. Apart from the effect natural environmental 
changes have (e.g. rain/sunny day) ephemeral factors such as vehicles or obstacles 
may induce multipath or other effects. Finally some protocols like
WiFi allow for transmission in different communication sub-channels 
according to conditions in the RF spectrum in the devices' vicinity,
for example if more than one device broadcasts in the same channel. 
As such, we also experiment with a more challenging and realistic scenario where the training data is
collected during one day and the testing data is from a different day, using a
smaller number of classes.  
The last two rows in Table~\ref{tab:time-results} contain the experiments used to estimate 
the effect of these conditions. When we train on one day and test using a different one,
all the aforementioned challenges are reflected in our results as even the RDCN
architecture obtains only a 10\% accuracy with the correct answer being in the
top five 30\% of the time. The real-valued networks
learn nothing as their accuracy is equivalent to random selection.
When we train on signals from two different days the network appears to learn and discard
the channel effects and create more robust signal features, since testing on a different day now achieves an increase in accuracy with the RDCN rising to 21\%. While
the other architectures improve as well, they remain considerably less accurate.

It is important to note that while we refer to Days 1,2,3
for ease of presentation, the data is not from three specific days but can span the 
entire 13 days. For example, the data used for one device might be from the first, 
sixth and eleventh day, while for a different device it might be from the second, 
third, and seventh day of the collection period. This allows us to truly stress test
our system under different environmental settings. In our opinion, this experiment highlights perhaps the toughest challenge in operating 
with raw I/Q data. Previous work that relied on protocol-specific techniques by extracting 
attributes (e.g., constellation phase-shift) obfuscated this obstacle since
sophisticated protocol-specific signal processing techniques can rectify these effects.
The fact that the RDCN network seems to be able to learn the channel
effects within such a complicated domain using signals captured from different
days is an encouraging result for the feasibility of protocol-agnostic fingerprinting
in realistic settings.

\subsection{Channel effect: signal-to-noise implications}

As the devices being monitored are likely to be in an external setting, the
quality of the signal can be affected by environmental factors resulting in
varying levels of noise. In this set of experiments we aim to investigate how
\system's performance is affected using datasets with different
signal-to-noise ratios. We train our models with 100 classes
using an equal number of training and testing samples (525)
that were captured over a mix of different days. We break down our data into
three different levels of SNR (\emph{Low} $\leq$2dB, 2dB$<$\emph{Med}$<$5dB and
\emph{High} $\geq$5dB). To make conditions more challenging we experiment with
testing and training combinations that have different SNR levels, as shown in
Table~\ref{tab:snr-results}. The complex-valued networks both achieve accuracy
higher than 98\% and our results show that they are unaffected by different
SNRs. The RDCN network achieves almost 100\% accuracy in all settings. The
real-valued networks perform relatively better in settings where the testing data is
of higher quality than the training but suffer when attempting to recognize signals
of \emph{lower} quality when trained on signals of \emph{higher} quality. The most indicative
such case is shown when the training used signals with \emph{High} SNR and
the testing occurred on \emph{Low} SNR with ANN and CNN both achieving an accuracy
of less than 27\%. We observe an interesting relation between this experiment and the
one comparing filtered and raw signals;
in both experiments the complex-valued networks appear
to be able to ``see through'' the noise and extract robust and meaningful
features whereas the real-valued networks are more susceptible to superficial
effects induced by noise.

\begin{table}[t]
\center
\caption{Accuracy when using different subsets 
of ADSB-1 with varying levels of SNR ratios.}
\begin{tabular}{ccccccc}
\toprule
\textbf{Train} & \textbf{Test} & \textbf{ANN} & \textbf{CNN} &\textbf{CDCN} & \textbf{RDCN} & \textbf{RDCN (top5)}\\
\toprule
\rowcolor{Gray}
Low & Med & 1\% &  98\% & 100\% & 100\% & 100\% \\ %Experiment 4F
Low & High & 75\% & 99\% & 99\% & 100\% & 100\% \\%Experiment 4E was 99 should be 75
\rowcolor{Gray}
Med & Low & 85\% & 85\% & 98\% & 99\% & 100\% \\%Experiment 4D
Med & High & 99\% & 100\% & 99\% & 100\% & 100\% \\%Experiment 4C
\rowcolor{Gray}
High & Low & 27\% & 26\% & 99\% & 99\% & 100\% \\%Experiment 4B
High & Med & 78\% & 78\% & 99\% & 100\% & 100\% \\%Experiment 4A
\bottomrule
\end{tabular}
\label{tab:snr-results}
\end{table}

\begin{figure}[t]
\centering
\includegraphics[width=0.8\columnwidth]{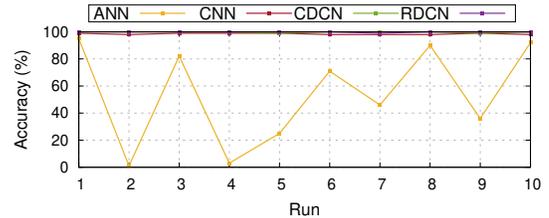}
\caption{Consistency of neural architectures over ten runs on the same
testing and training datasets.} 
\label{fig:deviation}
\end{figure}

\textbf{Model Reliability.} During this experiment we
encountered a discrepancy in the ANN model, shown in the first row of
Table~\ref{tab:snr-results}. To further investigate this,
we performed 10 runs for each architecture as shown in Figure~\ref{fig:deviation}.
We observe that the ANN network has a mean value of 54.1\% with a
standard deviation of 36.7 whereas the other networks remain consistent.
When the ANN performed poorly it
suffered from the first epoch, indicating that this issue stems from weight
initialization. The limitations of this simplistic, yet parameter-rich, model
are also evident when the number of devices increases and it completely fails
to scale, as discussed next.

\subsection{Class size effect}

\begin{figure}[t]
\centering
\includegraphics[width=0.8\columnwidth]{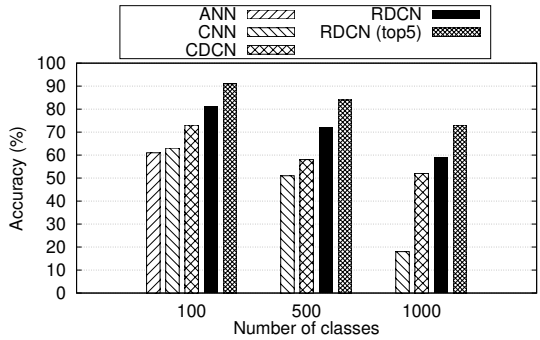} %experiments 1A, 1B, 2B
\caption{\system's accuracy for different number of classes.}
\label{fig:num_classes}

\end{figure}
We explore how our system fares when the number of classes increases.
As Figure~\ref{fig:num_classes} shows, we run three sets of
experiments where \system is required to differentiate between 100, 500, and
1000 sources. These experiments evaluate our system using an equal mix of
data from both the WiFi-1 and ADSB-1 datasets, in all setups
we use 218 signals for each device for training and 54 for testing.
One can easily observe that accuracy
diminishes as the number of classes increase. Interestingly it does so in a non-linear
fashion, with the RDCN achieving an accuracy of about 82\% for 100 devices and dropping 
to 72\% and 62\% for 500 and 1,000 classes respectively. It is also evident that 
the real-valued networks fail to produce models with sufficient descriptive power 
to follow the increase in cardinality, even though they have more parameters than the
CDCN architecture. Finally we observe a complete breakdown of the ANN
and a more rapid decline of the real-valued CNN when compared to its corresponding complex-valued
CDCN. The decrease in accuracy due to an increase in cardinality is known and expected~\cite{yang2017breaking}.

\subsection{Learning hardware imperfections} %after this switch to column RS6
If the data processing methodology is
not designed correctly then the models tend to latch onto (software-level)
identifiers as opposed to the hardware characteristics of the transmitting
device. To further demonstrate the feasibility of creating fingerprints
from the unique hardware imperfections that arise during a device's
manufacturing process, our next experiment aims to stress test our
system. Specifically, we leverage the \emph{WiFi-2} dataset obtained
from 19 identical smartphones flashed with the same firmware version,
configured with the same link-layer identifier (i.e., MAC address)
and transmitting the same data.
We present the results in Table~\ref{tab:identifier-results}.
To avoid latching on to either ephemeral or unique identifiers
we cut each signal in 6 parts and perform classification using one randomly selected segment. Each training sample comprises less than 17\% of the original signal, disrupting the continuity of the transmission, which leads to two primary benefits; it increases the diversity of the  dataset and guides the network towards finding features that identify the device based on the signal itself. Simply latching onto identifiers would lead to poor performance in this experiment.
In our current setup, our random segments are approximately 10$\mu$s 
worth of transmission or 1033 data points (out of the 6400 originally). This 
fragmentation prevents latching onto ephemeral or unique identifiers wherever 
they may be located in the signal, without requiring prior knowledge of where they may be located.
This allows our methodology to be applied to any protocol.

As can be seen in Table~\ref{tab:identifier-results}, this has a detrimental
effect on all the networks except RDCN, as its accuracy remains within 2-4\%
of when provided with the entire signal. This also shows that our cropping 
methodology prevents the models from using secondary protocol features (e.g., 
timestamps) as a discerning signal since, if that was the case, all the models
would continue to exhibit high accuracy.We believe that superiority of our model
is due to the existence of 
sufficient gated memory cells in the RDCN layer and its ability to learn features 
in \emph{context}. Once the signal is cropped and fed to the network its various
parts stem from different transmission phases (i.e., preamble, transient states, 
payload-data transmission, checksum etc). The morphology of the signal in these different 
phases differs drastically as seen in Fig~\ref{fig:signal_identifiers}, and the RDCN
network's ability to memorize features and work \emph{in context} is
the crucial difference to the other architectures that struggle with this diversity. This robustness, derived from capturing the essential transitional imperfections of the original transmitting device, will allow an authentication system to reject impersonating devices that have spoofed identifiers (e.g., MAC address) even if they are using stolen credentials or certificates.

\begin{table}[t]
\center
\caption{Accuracy when using random signal fragments to prevent
networks from learning software-level identifiers.}
\begin{tabular}{lccccccc}
\toprule
\textbf{Dataset} & \textbf{Classes} & \textbf{ANN} & \textbf{CNN} & \textbf{CDCN} & \textbf{RDCN} & \textbf{RDCN (t5)}\\ %experiment 5B,
\toprule
\rowcolor{Gray}
WiFi-2  & 19 & 21\% & 26\% & 27\% & 99\% & 100\%\\ %5B 
ADSB-1  & 100 & 1\%  & 15\% & 37\% & 98\% & 100\% \\ %4B 
\rowcolor{Gray}
WiFi-1  & 100 & 2\%  & 26\% & 24\% & 69\% & 90\% \\ %3B 
Mix     & 100 & 15\% & 34\% & 36\% & 80\% & 93\% \\%1C
\bottomrule
\end{tabular}
\label{tab:identifier-results}
\end{table}

\subsection{System Performance}

\begin{figure}[t]
\centering
\includegraphics[width=0.85\columnwidth]{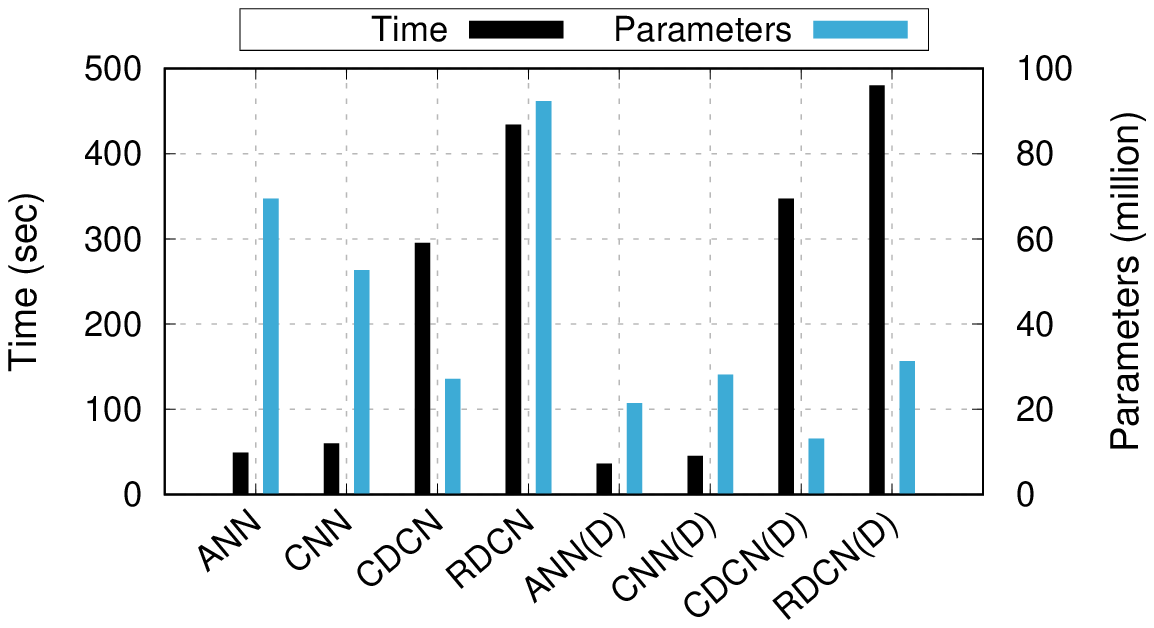}
\caption{Performance of different architectures. (D)
denotes the use of decimation with a factor of 2.}
\label{fig:performance}
\end{figure}

Figure~\ref{fig:performance} depicts
the processing time required to train a mixed dataset over a single epoch,
and the parameters contained by the model.
We find that the number of parameters does not always correlate
with the time required to process the data, as is evident for the ANN that has
almost double the parameters compared to CDCN yet takes less than a quarter
of the time for processing. This computational load, which
is independent of the number of parameters, exhibits an inverse correlation to
the accuracy of the networks.
We also include measurements from our decimation
experiment, in which the signals are decimated with a factor of two.
Decimation vastly reduces the number of parameters
and even though it leads to a signal half the size, %(6400 to 3200 points) 
the number of parameters is less than one third of the original.
RDCN is the ``heaviest'' model containing 30M parameters, down from 92.3M.
As we use all the decimated samples instead of just one,
we elevate decimation from a computation-reducing technique to
a data-augmentation strategy as well. While this improves accuracy,
it increases the processing time required by the complex-valued networks,
indicating the computational mismatch between the real and the complex-valued
counterparts.

\begin{figure}[t]
\centering
\includegraphics[width=0.9\columnwidth]{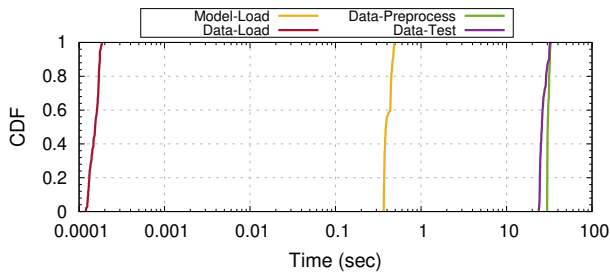}
    \caption{Time required for each testing subphase (RDCN).}
\label{fig:phases_cdf}
\end{figure}

Figure~\ref{fig:phases_cdf} plots \system's testing duration for each phase.
We use RDCN and conduct 100
runs on an idle server. Our dataset includes 11,200 signals from 100
classes, amounting to a 1.7G testing dataset. Here the
model has already been trained and we are interested in the time required for
the different phases required for testing. It is evident that computation is
dominated by the time required to pre-process and then test the data,
requiring between 23-33 and 29-33 seconds respectively. This experiment
reflects a scenario where the attacker monitors an area with 100 signal sources
and aims to classify them. In a scenario where \system is interacting with a
single device, e.g., for authenticating a device connecting to an access point,
verifying the device would require approximately 0.5 seconds %to authenticate the device
which is a reasonable requirement for bootstrapping authenticated communication between the two
endpoints. 

\section{Related Work}
\label{sec:related}

\textbf{Fingerprinting.} Numerous prior studies have proposed
approaches for fingerprinting devices, with browser fingerprinting attracting
much attention due to the ubiquitous presence of browsers~\cite{Eckersley:2010,MS12,CaoLW17,starov2017xhound,hupperich2015robustness,Acar:2013,englehardt2016online}.
Fingerprinting work, however, is not exclusive to browsers.
Kohno et al.~\cite{kohno2005remote} introduced the notion of remotely
fingerprinting a specific physical device over the network
by inspecting packet headers and identifying the clock skew due to the hardware
characteristics of a given device. Several subsequent studies have explored 
alternative approaches for detecting clock skew or other protocol-specific 
features~\cite{arackaparambil2010reliability,jana2009fast,miettinen2017iot,
radhakrishnan2014gtid,barbosa2013flow}. Brik et
al.~\cite{brik2008wireless} demonstrated that
hardware imperfections can be leveraged as successful fingerprinting 
features;
however their approach relied on protocol-specific information
since during the last steps of demodulation one needs to know the utilized 
modulation scheme in order to estimate the difference from the ideal 
expected symbol constellation. Such approaches, while accurate,
rely on protocol-specific noticeable imperfections that 
need to be \emph{manually} identified and extracted using handcrafted techniques
which is problematic for unknown protocols.
It is also important to note that
their data was collected from an indoor testbed
with statically placed
transmitters -- in practice a fingerprinting system would face
significantly harsher conditions that affect performance, as demonstrated
by our experiments. Also, their data was collected using one receiver which
can prove problematic in realistic settings (see Section~\ref{sec:future}). Such limitations are common in
this line of research (e.g.~\cite{Vo-Huu:2016}).
Dey et al.~\cite{dey2014accelprint} argued that the physical imperfections of
smartphones' accelerometer sensors rendered devices fingerprintable, as do the audio 
components~\cite{das2014you}.

In contrast to prior work we propose and develop a widely applicable fingerprinting 
approach, which presents a paradigm shift as it removes many assumptions and practical 
limitations that affected prior techniques. Specifically our work differs across multiple
dimensions. First, our system is fully passive and does not require any form of interaction with the user or the target device. Second, our approach is completely protocol-agnostic and 
does not require any information about the underlying protocols, operating system or 
applications running. As a result, our proposed system is not affected by differences in software
implementations across platforms or different versions, changes to the device's software (e.g., updates),
or the use of custom protocols. Moreover, we present an extensive experimentation that 
explores the effect of multiple dimensions of the data collection and model creation process.
Finally, our system design guides the neural network towards learning the effect of the 
device's hardware on the transmitted signal, thus overcoming obstacles presented by 
software-level defenses.

\textbf{Complex-valued networks.} Trabelsi et al.~\cite{trabelsi2017deep}
laid the theoretical foundations for using complex-valued
networks and proposed appropriate building blocks such as complex
batch normalization and complex weight initialization to facilitate training
with complex data. Wolter et al.~\cite{wolter2018gated} subsequently
introduced and successfully trained a complex-enabled GRU and showed that it
exhibits both stability and competitiveness.

\textbf{Deep learning in RF.}
Recent papers~\cite{ye2018power,o2017introduction,rajendran2018deep}
highlighted the potential benefits of using deep learning
for signal processing as well as tackling fundamental problems like channel estimation~\cite{ye2018power,rajendran2018deep},
modulation classification~\cite{o2017introduction} even directly recovering symbols without
demodulation~\cite{ye2018power}; all these networks rely on real valued networks and treat the
naturally complex data as two real valued channels. The authors
of~\cite{o2017introduction,rajendran2018deep} stated
the lack of complex-valued networks as an open research question and highlighted the
mismatch between the common use of complex numbers in almost all signal
processing algorithms and the real-valued nature of the networks.
O'Shea and Hoydis~\cite{o2017introduction} identified some crucial
problems such as the non-holomorphic nature of common activation
functions, which was addressed in~\cite{trabelsi2017deep}.

In an independent concurrent study, Restuccia et al.~\cite{DBLP:journals/corr/abs-1904-07623}
have proposed the use of deep learning for fingerprinting devices based on
RF signals. However, their study presents multiple significant limitations
compared to our work. First, their system relies on protocol-specific processing
of the signal, which introduces an important deployment constraint compared to 
our protocol-agnostic design.
Next, their system works with real-valued data,
as opposed to our complex-valued data approach, which inherently results in a loss
of information during the down-converting process. Moreover, their system 
is built on top of a CNN which, as demonstrated by our experimental evaluation,
is considerably less accurate -- our RDCN architecture outperformed the CNN
across all experiments. Perhaps the most critical limitation is that they provide the 
entire signal for training and testing; as we highlighted in our
experiments, that guides the CNN towards learning the device identifiers (e.g.,
MAC address or aircraft ID) as opposed to actual hardware imperfections.
Next, the majority of their experiments is conducted with a very small number of 
classes; their largest experiment includes numbers for up to 500 classes but, as shown in our Figure~\ref{fig:num_classes},
the CNN architecture's accuracy deteriorates significantly for more devices.
Finally, they train their WiFi and ADS-B networks \emph{separately}, whereas we experiment
with multi-protocol systems where the networks are fed both types of signals without
any prior knowledge of the underlying protocols. Overall, while their work presents
similarities with ours, we believe that our paper presents significant advantages
for practical deployment, while also demonstrating superior accuracy under more challenging
scenarios. We also explore the effect of many practical dimensions of the 
data collection and network training processes that are
omitted in their work.

\section{Discussion and Future Work}
\label{sec:future}

\textbf{Deployment.} Our techniques can be deployed in a
defensive capacity in multiple scenarios ranging
from passively authenticating devices~\cite{kroon2009steady,talbot2003specific,riezenman2000cellular,scanlon2010feature}
to detecting unauthorized devices in restricted areas~\cite{bisio2018unauthorized}
or unmasking spoofed transmissions (e.g., 
``wormhole'' attacks~\cite{hu2003packet}). Another potential
application is detecting the illegitimate use of credentials/certificates from unauthorized transmitters.

\textbf{Agnostic vs generalizable techniques.} 
An interesting related line of work explores using a range of mathematical and statistical tools that are broadly applicable across protocols, such as the work Danev et al.~\cite{danev2009transient,danev2012physical} or Klein et al.~\cite{klein2009application}. The main difference between our work 
and the above is that we explore the feasibility of fingerprinting  
\emph{without knowledge of the modulation or the carrier frequencies}.
While our system's accuracy does indeed benefit from doing generic preprocessing, our system works in a completely agnostic manner given a capture of signals within a given frequency range; it also handles devices that transmit at multiple different frequency ranges,
without any ``guiding'' information provided to the system.
In contrast while these prior methods are indeed generalizable and applicable across protocols, they \emph{require} the ability to locate the signal and move it to an IF while filtering and clearing the signal sufficiently to extract the statistical or mathematical properties used in defining the fingerprint. 

\textbf{Hardware-based bypassing.}
While our technique is inherently robust to OS and software level attempts of subversion, 
attackers could try to bypass it using
hardware-based approaches and trying to mimic
the physical characteristics that were learned by our network for a specific device.
This presents several significant challenges in practice,
and would require considerable effort and money~\cite{brik2008wireless}, if at all feasible.

Danev et al.\cite{danev2010attacks} explored two attacks for circumventing physical authentication systems: (i) a feature replication scheme for bypassing modulation-based schemes, and (ii) using a high quality signal generator to faithfully reconstruct a captured signal and deceive low-level physical authentication schemes such as transient-based approaches. The first attack is ineffective against our system since modulation features are irrelevant; the second, more powerful attack could potentially trick our system within lab conditions. In their evaluation they captured and replicated signals in a lab environment with devices placed on a tripod 30cm away from the receiver. In a realistic scenario, capturing a signal in a given location would consist of transmissions that are affected by the given channel and would be re-transmitted in the channel of the new location, which would adversely affect the re-transmitted signal. The spectrum conditions at the new location might also affect the re-transmitted signal in ways that the authors did not explore. Consider  capturing a WiFi signal at channel 1; when re-transmitting at the relay location prevalent spectrum conditions may contain other WiFi devices transmitting at channel 1, thus, inducing conflicts. Therefore, while relay transmission from high-end quality hardware 
could pose a threat to our fingerprinting scheme, significant experimentation is required
for exploring the feasibility of such an attack under realistic non-lab conditions.
As such, we argue that our proposed techniques present a robust
augmentation to existing authentication (and other defensive) schemes.

\textbf{Scalability.}
The limitations on softmax-based classification,
also referred to as the ``softmax bottleneck'', have been studied extensively
in the work of Yang et al.~\cite{yang2017breaking}. Softmax adaptations like
Sigsoftmax~\cite{kanai2018sigsoftmax} might allow for higher cardinalities.
Discriminatory schemes such as
Siamese networks~\cite{bromley1994signature} can facilitate an open set classification strategy,
providing robustness in the face of a large number of classes.
Our RDCN and CDCN networks can be retrofitted to power
a Siamese architecture with the appropriate loss function; finding the appropriate 
loss metric for the comparison of RF complex data is a challenging and interesting problem.

\textbf{Signal types.} In our experimental evaluation we used
WiFi and ADS-B signals collected both in the wild and in laboratory settings.
Common user devices also use other signal families and protocols 
such as GSM and 4G. As such, in the future we plan to experiment
with other prevalent signal types; however, given that \system works on raw
data and is agnostic of the underlying protocols we believe that our techniques
can be directly applied to a wide variety of wireless signals.

\textbf{Model transferability.} Similarly to how the transmitters
leave a unique fingerprint on signals, which is the motivating observation
behind \system, the receiving hardware may also uniquely affect the signal.
As such, in scenarios where the training and testing data have been collected by
different devices 
a model's accuracy could
be impacted. In the datasets used in our experiments the testing and training data
was collected by multiple receivers and antennas.
However, as the datasets
do not have labels regarding the capturing device/antenna, we were unable to
quantify this effect. 
We consider the in-depth exploration of the transferability of models across
different monitoring devices as part of our future work. Given our current
results we believe that leveraging multi-antenna or multi-device monitoring
setups, where signals are collected from several receivers with different
polarizations, can lead to \system learning the imperfections of the
transmitting devices while ``ignoring'' the receiving hardware's effect. 

\textbf{Model interpretability.} This study presents the first exploration
on the feasibility of using deep complex models for fingerprinting devices
based on their raw RF transmissions. Our extensive experimental evaluation
demonstrates that such techniques are well suited for such tasks. While we
have designed our novel models based on specific domain-based observations 
and well-defined design goals, we consider an in-depth exploration of these 
network architectures that deconstructs and interprets their predictions 
(e.g., using LRP~\cite{lrp}) an important aspect of our future work.

\section{Conclusion}
\label{sec:conclusion}
In this paper we demonstrated the effectiveness of novel complex-valued networks,
as they can produce powerful models able to express fingerprints that uniquely identify
devices based on the wireless signals they transmit. Specifically, we presented a system 
that can identify artifacts ingrained in the wireless signals due to the imperfections
inherent in each transmitter's circuitry.
One of our main motivations was to explore the feasibility of building 
an RF-based fingerprinting system that can act as an additional factor of authentication,
able to defend against powerful attacks such as impersonation, 
digital certificate theft, or relay attacks. 
Our prototype demonstrates that such a system can be readily deployed 
and operate within realistic environments and without constraints on
the type of signal that devices need to transmit, knowledge of the underlying protocol 
or software implementations. Through our extensive experimental 
evaluation we highlighted the challenging nature of this task, and quantified the effect 
of different factors across multiple dimensions of the testing and training processes.
While our approach is robust and effective under realistic conditions, we believe that 
our work is an important exploratory first step within a vast and challenging new space,
and has identified several interesting future directions.

\bibliographystyle{IEEEtran}
\bibliography{paper}

\appendix
\label{sec:app}

\subsection{Background}

\subsubsection{General training and backpropagation.} In the general case, we use a
subset of available data as training samples that the networks attempt to best
describe by adjusting their parameters accordingly. This adjustment step is
usually done by passing a number of samples trough the network, acquiring the
output and combining it with a function that is usually referred to as a loss
function or cost module. This loss function outputs a value, usually scalar,
that is a metric representing the error or distance of the network's output
relative to the ground truth. Driving the loss towards zero is a desirable
goal, as it tends to lead to a better performing network; this is done by
adjusting the parameters of the network in such a way that if we passed the
same sample we used to obtain the loss for this iteration the new output's
loss will be smaller. This is generally performed with a method known as
\emph{backpropagation} (proven to be able to train deep 
networks~\cite{werbos1974beyond}), which essentially uses the chain differentiation rule
to discover the contribution of each of the network's parameters to the
computed loss. The algorithm starts computing the expressions from the final
layers of the network (closest to the output), towards the start, computing the
error contributions of each layer's representation based on the previous
layer's representation of the input. The general algorithm employed for the
actual adjustment step, i.e, the magnitude and direction of change, is given by a
gradient descent-based algorithm such as Stochastic Gradient Descent (SGD) or
variants like Adam~\cite{kingma2014adam} and AdaDelta~\cite{zeiler2012adadelta}.

\subsubsection{SigMF: storing I/Q data.}
\label{sub:sigmfdata}
Storing and sharing RF data had been a long standing problem.
Following the DARPA RF initiative, the SigMF specification was established in 2017 to
tackle this issue and allow for a flexible yet well-specified standard
for sharing RF data in an efficient manner.
The raw I/Q data is specified in an interleaved form in a binary file where 
the precise location of each signal in the data file is denoted in an 
accompanying metadata file in a JSON format.
The metadata files also contain a host of additional information such as capture details, 
sampling rate and may optionally be extended to support new custom headers.
All the datasets used in this work utilize the SigMF~\cite{sigmf} specification.

\subsubsection{Quadrature Modulation}
\begin{figure}[th]
\hspace{2cm}
\includegraphics[width=0.5\columnwidth]{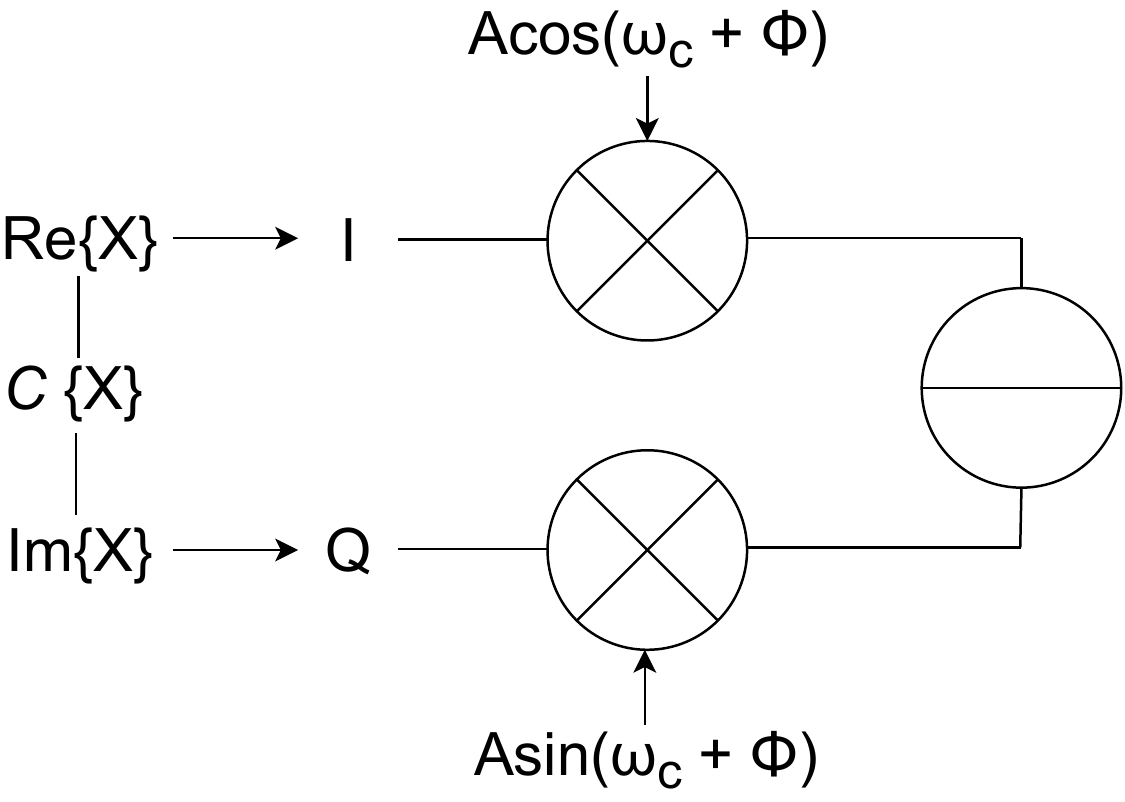}
\caption{Quadrature Modulation Overview and I/Q data. An inverse procedure is
followed on the receiver side in order to obtain the I/Q pair from the
modulated signal.}
\label{fig:quadrature_modulation}
\end{figure}

Figure~\ref{fig:quadrature_modulation} depicts a schematic of an ideal 
quadrature modulator. The output signal is composed of a linear combination 
of signals in-phase or ``I'' component and the quadrature or ``Q'' component.
Besides its apparent simplicity, quadrature modulators can synthesize \textbf{any} 
real signal.

\subsubsection{Inverse Square Root of Matrix V}
\label{theorem:tikonov}
To guarantee the existence of the inverse square root we
leverage the Tikhonov regularization~\cite{tikhonov2013numerical}
that is performed by: $\epsilon$ \textbf{I} +
\textbf{V}. Mean subtraction and multiplication by the inverse square root of the
variance guarantees a standard complex distribution with mean $\mu=0$,
covariance $\Gamma=1$ and relation $C=0$. As in the real case batch
normalization procedure, two learnable parameters are used: the shift parameter
$\beta$ and scale parameter $\gamma$. These translate and scale a layer's
input along new learned principal components to achieve the desired
variance.

\subsection{Data Augmentation using Rotation, Scaling and Noise}
\label{app:data_augmentation}

While our experiments using decimation revealed that it can have a positive impact,
we found that for our particular task these transformations actually have a negative 
impact on our system's performance. Below we provide more details for these augmentations.

\subsubsection{Rotation and scaling.}
Similarly to their 2D counterparts (images), signals can be rotated (phase
shift) and scaled (magnitude scaling). This lead to an expansion of the amount
of available data by introducing reversible transformations.

We induce rotation by adding a phase shift and scaling
by altering the magnitude of each complex data point. To calculate the phase
and magnitude we convert each point to polar form, perform the necessary calculations,
and recover the I/Q components using Euler's equations. Formulae~\ref{eqn:rotation}
describe the procedure for rotation, scaling follows the same principle.
\begin{equation}
    \label{eqn:rotation}
\begin{aligned}
    \begin{split}
    &Original\, Signal= I + Qj \\
    &Polar\, Form = \rho e^{\phi}, \rho = \sqrt{I^2 + Q^2}, \phi = \arctan{(I,Q)} \\
    &Rotation(\theta) = \rho e^{\iota \phi} e^{\iota \theta}  = \rho e^{\iota (\phi + \theta)} \\
    &I'= \rho \cos{(\phi + \theta)} , Q'= \rho \sin{(\phi +\theta)}j , \\
    &Rotated \, Signal=I'+Q'j
    \end{split}
\end{aligned}
\end{equation}

\subsubsection{Adding noise - channel effects.}
Perhaps the most prominent problem regarding RF data is the effect the channel has 
to captured signals under different channel conditions.
As shown in Figure~\ref{fig:signal_overview} signals from the same device captured
over different days can differ drastically. The fundamental way the signal
mutates under different channel conditions introduces a mismatch between the
data used for training and signals captured for testing at a later time when
the ambient channel conditions deviate significantly between them.
We evaluated two methodologies and their efficacy in improving the model's generalization
in regards with being able to classify signals from a day not part of the training set:
(i) Augmenting the dataset by introducing samples from different days in the training data 
(ii) Augmenting captured signals synthetically by introducing Rician or Rayleigh fading
using a SISO channel model, based on~\cite{taranali}.

\subsection{Hyperparameters}
\label{app:hyperparameters}

The hyperparameters used for each architecture are
obtained using Spearmint~\cite{snoek2012practical}, thus
allowing 500 runs for each (complex) architecture. The input data
used in Spearmint for optimizing the hyperparameters is a balanced dataset that
It is important to note that even though using Spearmint and its underlying
Bayesian optimization engine requires less tries than conventional searches such as 
grid search, we were still unable to exhaustively explore all options as that would require a prohibitive 
amount of time and used the best values after 500 tests.
We included the following hyperparameters in the Spearmint exploration for each architecture:

\emph{CDCN}: Learning rate, kernel/stride/padding of the convolutional layer,
the size of the output dense layer. 

\emph{RDCN}: Learning rate, weight decay, the timestep ($\mu$s) used by the sequencer 
to split the signal,the size of the hidden layer of the LSTM,
the number of layers in the LSTM unit and whether it should be bidirectional or not.

For both architectures the Spearmint search configurations also include the factor
of decimation factor $M$ and explore whether the following layers 
should be part of the architecture or not:
(i) batch normalization for the input layer,(ii) batch normalization for each layer independently.
For the decimation factor our exploration range lies between two and four, given our 
sampling rate factors above 2 violate Nyquist criterion, and the original signal (for the WiFi category) 
can no longer be faithfully reconstructed. Since our goal in this work is rather to fingerprint 
the devices and \emph{not} to reconstruct the original signal we evaluated the impact 
of using higher decimation factors, even if that would violate Nyquist's theorem. We identified 
that indeed the best results where obtained with a decimation factor of 2 (which obeys the 
theorem). As a final note, while the non-conforming ($M>2$) decimation factors 
produced worst results they also led to less computations and less parameters 
which in some applications might be a wanted outcome.

\textbf{Optimized hyperparameters.}
Our exploration revealed the following hyperparameters and network configurations:
\emph{CDCN:}
lr=1e-4, weight decay=9e-5, kernel size=32, stride=1, padding=1, dense layer=4096.

\emph{RDCN} lr=1e-4, weight decay=9e-5, single layer unidirectional LSTM with 1024 hidden units 
and sequencer time step set to $1\mu$ leading to 64 vectors of 100 data points for the baseline setting.

For both networks the batch normalization layers were found to be of critical importance and
the best performing activation unit was found to be $\mathbb{C}$ReLU. We believe that since 
our channels are mixed channels containing information that is not entirely real or
imaginary, the effect of ZReLU that constricts activations to the first quadrant is 
over-constraining the network, whereas the outcome of $\mathbb{C}$ReLU that allows activation 
on positive values for both channels shows better results.

It is important to note that these hyperparameters are optimized for our design goals, i.e., operating on
raw I/Q data and with minimal pre-processing being performed. Alternate data
representations or extra pre-processing steps would potentially render these hyperparameters 
invalid and a re-exploration using Spearmint (or a conventional, albeit slower,
grid search) to obtain the optimal parameters would be required.

\end{document}